\RequirePackage{fix-cm}
\documentclass{svjour3}                     % onecolumn (standard format)
\smartqed  % flush right qed marks, e.g. at end of proof
\usepackage{graphicx}
\usepackage[T1]{fontenc}
\usepackage[utf8,latin1]{inputenc}
\usepackage{epsfig}
\usepackage{amssymb}
\usepackage{natbib}        
\usepackage{xspace}
\usepackage{rotating}
\usepackage{amsmath}
\usepackage{scrextend}
\usepackage{url}

\usepackage{breakurl}
\usepackage[hidelinks]{hyperref}

\usepackage{url}
\usepackage{xcolor}
\usepackage{bm}
\usepackage{sgl-commands}

 \journalname{my journal}

\begin{document}

\title{Searching for strong gravitational lenses}

%\subtitle{}


\author{Cameron Lemon$^{1}$, Fr\'ed\'eric Courbin$^{1}$, Anupreeta More$^{2, 3}$, Paul Schechter$^{4}$, Raoul Ca\~{n}ameras$^{5, 6}$, Ludovic Delchambre$^{7}$, Calvin Leung$^{4}$, Yiping Shu$^{5, 8}$, Chiara Spiniello$^{9, 10}$, Yashar Hezaveh$^{11, 12}$, Jonas Kl\"{u}ter$^{13}$, Richard McMahon$^{14, 15}$}

\institute{\email{cameron.lemon@epfl.ch}
%\and 
\\
$^1$Institute of Physics, Laboratory of Astrophysics, Ecole Polytechnique F\'{e}d\'{e}rale de Lausanne (EPFL), Observatoire de Sauverny, 1290 Versoix, Switzerland
\\
$^{2}$The Inter-University Centre for Astronomy and Astrophysics (IUCAA), Post Bag 4, Ganeshkhind, Pune 411007, India
\\
$^{3}$Kavli Institute for the Physics and Mathematics of the Universe (IPMU), 5-1-5 Kashiwanoha, Kashiwa-shi, Chiba 277-8583, Japan
\\
$^{4}$MIT Kavli Institute 37-635, 77 Massachusetts Avenue, Cambridge, MA, 02138-4307, USA
\\
$^{5}$Max-Planck-Institut fur Astrophysik, Karl- Schwarzschild-Str. 1, 85748 Garching, Germany
\\
$^{6}$Technical University of Munich, TUM School of Natural Sciences, Department of Physics, James-Franck-Stra{\ss}e 1, 85748, Garching, Germany
\\
$^{7}$Institut d'Astrophysique et de G\'{e}ophysique, Universit\'{e} de Li\`{e}ge, 19c, All\'{e}e du 6 Ao\^{u}t, 4000, Li\`{e}ge, Belgium
\\
$^{8}$Purple Mountain Observatory, Chinese Academy of Sciences, 3R8H+9RC, Tianwentai Rd, Xuanwu, Nanjing, 210023, PR China
\\
$^{9}$INAF - Osservatorio Astronomico di Capodimonte, Via Moiariello 16, I-80131 Naples, Italy
\\
$^{10}$Sub-Department of Astrophysics, Department of Physics, University of Oxford, Denys Wilkinson Building, Keble Road, Oxford OX1 3RH, UK
\\
...continued on page 2...
}

%Tom Collett, Christine Ducourant, Alberto Krone-Martins, Keren Sharon, Rachel Webster

%\institute{N. Name \at
%              Affiliation\\
%              Address \\
%		   City, Country\\
%              \email{name@somewhere.edu}           
% }

%\date{Received: date / Accepted: date }

%\input{xxx}
\authorrunning{Lemon et al.}
\titlerunning{Searching for lenses}
\maketitle

\begin{abstract}
Strong gravitational lenses provide unique laboratories for cosmological and astrophysical investigations, but they must first be discovered -- a task that can be met with significant contamination by other astrophysical objects and asterisms. Here we review strong lens searches, covering various sources (quasars, galaxies, supernovae, FRBs, GRBs, and GWs), lenses (early- and late-type galaxies, groups, and clusters), datasets (imaging, spectra, and lightcurves), and wavelengths. We first present the physical characteristics of the lens and source populations, highlighting relevant details for constructing targeted searches. Search techniques are described based on the main lensing feature that is required for the technique to work, namely one of: (i) an associated magnification, (ii) multiple spatially-resolved images, (iii) multiple redshifts, or (iv) a non-zero time delay between images. To use the current lens samples for science, and for the design of future searches, we list several selection biases that exist due to these discovery techniques. We conclude by discussing the future of lens searches in upcoming surveys and the new population of lenses that will be discovered.

%Serendiptious discoveries until 1990s -> first successful targeted searches in the radio and from SDSS spectra -> samples of >100 lenses

%Now all-sky surveys at all wavelengths are pushing deeper and higher-resolution, with individual searches finding up to 100s of new lens candidates.

%Brief description of the physical characteristics of the deflector and source populations.

%Various algorithms can be used for many lens types in many datasets; we split our discussion into searches which often target one of four unique lensing features: magnification, multiple images, multiple redshifts, and time delays.

%We list several biases from these search methods that can exist

%We finish by discussing the future of lens searches focussing on upcoming surveys and the new population of lenses that will be discovered.

% add further keywords with "\and" 
\end{abstract}

% LD: I'm adding a table of contents here, just to have a global overview. We should later remove it.
\setcounter{tocdepth}{3}
\tableofcontents

%%%%%%%%

\newcommand\blfootnote[1]{%
  \begingroup
  \renewcommand\thefootnote{}\footnote{#1}%
  \addtocounter{footnote}{-1}%
  \endgroup
}
\deffootnote{0em}{0em}{\thefootnotemark\quad}

\blfootnote{\noindent $^{11}$D\'{e}partement de Physique, Universit\'{e} de Montr\'{e}al, Montreal, QC H3T 1J4, Canada
\\
$^{12}$Center for Computational Astrophysics, Flatiron Institute, 162 Fifth Avenue, New York, NY 10010, USA
\\
$^{13}$Department of Physics and Astronomy, Louisiana State University, 202 Nicholson Hall, Baton Rouge, LA 70803, USA
\\
$^{14}$Institute of Astronomy, University of Cambridge, Madingley Road, Cambridge CB3 0HA, UK
\\
$^{15}$Kavli Institute for Cosmology, University of Cambridge, Madingley Road, Cambridge CB3 0HA, UK
}

\label{chapter9}
\section{Introduction}
\label{sec1:intro}

The first measurement of the deflection of light by a gravitational field was the angular displacement of stars close to the Solar limb. The experiment, carried out during the 7-minute-long Solar eclipse visible on 29 May 1919, measured a small displacement of stars in the Taurus constellation, of the order of 1.75 arcseconds \citep{Dyson1920}, supporting the factor of two increase in the deflection angle prediction of general relativity over Newtonian gravity. Should the Sun be counted as the first lens "discovery"? Throughout this chapter we will be concerned with strong gravitational lensing, in which multiple images of the source are observed. Though the 1919 experiment would not count as the first strong lens, it is the first genuine observation of the gravitational lensing phenomenon. % and one of the most beautiful illustrations of how astronomical observations can serve fundamental physics. 
The first strong lens discovery would wait 60 years.

The relevant calculations of multiple imaging were first published by \citet{chwolson1924}, but a flurry of work was sparked a decade later by Einstein's 1936 publication on the subject \citep{einstein1936}. In fact, Einstein had made similar unpublished calculations as early as 1912, possibly to explain the intensity of a newly discovered Nova, Nova Geminorum 1912, but eventually concluded that the low probability, symmetrical lightcurve, and achromaticity of lensing were an incompatible explanation \citep{sauer2008}. As Einstein had only considered stars as both the lens and source, he correctly concluded that such image separations were unobservable. In 1937, Zwicky noted that galaxies or clusters are much more likely lenses to provide observable lens systems \citep{zwicky1937}. After the discovery of quasars in 1963, they were quickly recognised as a possible variable lensed source by Refsdal for measuring the Hubble constant, alongside supernovae \citep{refsdal1964}. Indeed, the first lens discovered in 1979 was a doubly imaged lensed quasar, QJ0957+561, a.k.a. ``The Twin Quasar'' \citep{walsh1979}.

The 1980s saw a few more serendipitous discoveries, including the first quadruply imaged lensed quasars: PG1115+080 \citep{weymannn1980} and Q2237+030 \citep[a.k.a. the Einstein Cross,][]{huchra1985}. %These searches relied on the magnification bias of lensing, namely that for the apparently brightest sources, an increased fraction will be lensed. 
The same decade saw the discovery of giant luminous arcs in the galaxy clusters Abell 370 and 2244--02 \citep{lynds1986, soucail1987}. The arcs were soon spectroscopically confirmed to be behind the cluster \citep{soucail1988, lynds1989}, confirming Zwicky's prescient suggestion that clusters were likely gravitational lenses. %, and thus the first cluster gravitational lenses as suggested by \citet{paczynski1987}.

%The first systematic searches for strong lensing at galaxy scale were carried out both in the optical and radio domains. In both cases, the search method took advantage of the magnification bias caused by lensing on the luminosity function of the lensed sources at large distances such as quasars and radio sources. To summarize the effect of magnification bias on a sample of lensed sources: lensed sources are seen brighter because they are lensed and therefore magnified. %CS: I will personally move the "To summarize...magnified" to a footnote
%As detailed in Chapter~1, this effect depends on the luminosity function of the population of sources prior to lensing. The steeper the luminosity function, the higher is the lensing probability. Some of the first lens searches therefore selected apparently bright sources among objects with steep luminosity functions such as quasars and radio sources and then organised observation campaigns to obtain high spatial resolution imaging in the hope to detect multiple images, Einstein rings or arcs produced by massive and compact objects at lower redshift on the line of sight. These searches are often called "source-selected", because the sample pre-selection is done on the properties of the background, lensed source. %a distant source. 
Systematic lens searches soon began, as high-resolution imaging with the \textit{Hubble Space Telescope} (\textit{HST}) and at radio wavelengths was used to look for multiple imaging of quasars. The first successful search in the optical was the doubly imaged quasar UM~673 \citep{Surdej1987}, quickly followed by the quadruply imaged quasar H1413$+$117 a.k.a. the ``Cloverleaf'' \citep{Magain1988}. In the radio, systematic searches found the first full Einstein ring \citep{Hewitt1988} and multiply-imaged radio sources such as MG~2016$+$112 \citep{Lawrence1984, Narasimha1984}. By the early 2000s, the Cosmic Lens All-Sky Survey \citep[CLASS;][]{myers2003} found 22 lenses within a sample of 13,500 radio sources.  These searches are often called ``source-selected'', because the sample pre-selection is made on the properties of the background, lensed source.

One difficulty with source-selected lens candidates resides in their confirmation as real lenses: the glare of the bright quasar images masks the often faint lensing galaxy. The task is particularly challenging with ground-based data, although image deconvolution techniques have been effective for deblending in optical imaging \citep[e.g.][]{Courbin1998, Burud1998} and spectroscopy \citep[e.g.][]{Eigenbrod2007}.
High-resolution data from \textit{HST} or from ground-based Adaptive Optics (AO) remains the best way to confirm lenses. %They are also the main sources of information to constrain lens models. 
%CS: could we add few references here? 
%Examples of large systematic searches for galaxy-quasar lensing in the optical include the SDSS lens survey \citep{oguri2006, inada2012} and more recent work using a combination of Gaia astrometry and a mix of complementary imaging surveys \citep{lemon2017, lemon2018}.
Source-selected lens samples are mainly limited to quasars, but there are $\sim$1000 times more galaxies than quasars and therefore about as many more galaxy-galaxy strong lenses than galaxy-quasar strong lenses. In the former, the lens galaxy light typically dominates, even taking into account the magnification of the source. Searches for galaxy-galaxy lenses have thus focused on a pre-selection of galaxies potentially acting as lenses and are therefore called ``lens-selected''.

Cluster lenses are also typically lens-selected, as deeper imaging is obtained for known clusters to search for multiple images. The clusters are first found through their X-ray emission \citep[e.g.,][]{ebeling2001}, or optical imaging surveys \citep[e.g.][]{gladders2005}, and subsequent \textit{HST} follow-up reveals, on average, at least 1 giant arc per cluster \citep{horesh2010} for the most massive (X-ray-bright) clusters, let alone many more compact sets of multiple images. While the discovery of multiple images in any cluster is very likely, it is a question of the depth and resolution of available imaging, as indeed deep observations of the most massive isolated galaxies can also often uncover faint background lensed sources \citep[e.g.,][]{collett2020}.

For both lens- and source- selected searches, wide-field imaging and spectroscopic surveys at various wavelengths are now the main catalyst for developing new searches, as the depth and quality of the data allows not only for many more lenses to be discovered, but also new types of lenses. The Sloan Digital Sky Survey (SDSS) was the first such survey, with targeted campaigns yielding new lensed galaxies from imaging \citep[e.g.,][]{belokurov2007} and spectra \citep[e.g.,][]{Bolton2004}, and lensed quasars from a combination of spectra and imaging \citep[e.g.,][]{pindor2004, oguri2006}. As modern surveys have ever increasing depth, area, and quality, both machine learning and citizen science have played a pivotal role in dealing with the enormous number of candidates from the increased depth and area \citep[e.g.,][]{marshall2016, Jacobs2019, Rojas2021}. Entirely unique datasets, such as the high-resolution catalogue of \textit{Gaia}, have allowed for the efficient identification of bright lensed quasars across the whole sky \citep[e.g.,][]{stern2021, lemon2023}.

The aim of this chapter is to present a detailed review of the various search methods presented in the literature for finding strong gravitational lenses of all source and lens types, and all image separations. 
The above dichotomy between source- and lens-selected samples often represents only the first filtering of a survey's catalogue; the subsequent selection algorithms are where most of the filtering happens. Data types, wavelengths, and source types are all possible ways to subdivide the many lens searches in the literature. We find it more helpful, however, to avoid repetition by broadly sorting the search methods into those targeting a common lensing feature, that is to say, either a magnification, multiple imaging, multiple redshifts, or an associated time delay. 

This chapter is outlined as follows: in Sections \ref{sec2:classes_deflectors} and \ref{sec3:classes_source} we describe the properties of the lens and source populations respectively. In Section \ref{sec4:seclection_methods}, we provide an overview of the literature in terms of selection methods following the categories described above. Possible biases and selection effects from these searches are described in Section \ref{sec5:bias}, and in Section \ref{sec6:future} we discuss how lens finding may evolve in the coming decade, in a context where key facilities such as \textit{Euclid}, Rubin-LSST, and Roman, will be delivering actual data.

%This reflects the current "split" in the lensing community between those more interested in the study of lensed sources and those who focus either on the properties of foreground deflectors or on the cosmological applications of strong lensing such as time-delay cosmography (e.g., \citealt{Suyu17}).
%The difficulty of the enterprise that we try to describe in this chapter, resides in devising techniques using individually or collectively all datasets available to us. These include systematic imaging and spectroscopic surveys in the optical, in the near-IR and in the radio, sometimes along with astrometric measurements, and in the future with time domain information. Successful techniques must make subtle use of the specific characteristics of the lenses, of the sources, all in combination with the impact of lensing on the sources. 

%The present chapter emphasizes which of the lens and source properties can be taken to our advantage to maximize the efficiency of lens searches. We describe selection methods of samples of potential lensing systems and the associated mathematical and statistical techniques used to dig into these samples, e.g., machine learning, citizen science. We finally describe possible avenues to evaluate the selection functions of lens finding algorithms, probably the most challenging task in any lens-finding activity, and we conclude on how lens finding may evolve in the coming decade, in a context where key facilities such as Euclid, Rubin-LSST, Roman, SKA, will be delivering actual data.  

%%%%%%%%
\section{Deflector classes}
\label{sec2:classes_deflectors}

\begin{figure*}[htbp]
    \centering
    \includegraphics[width=0.98\textwidth]{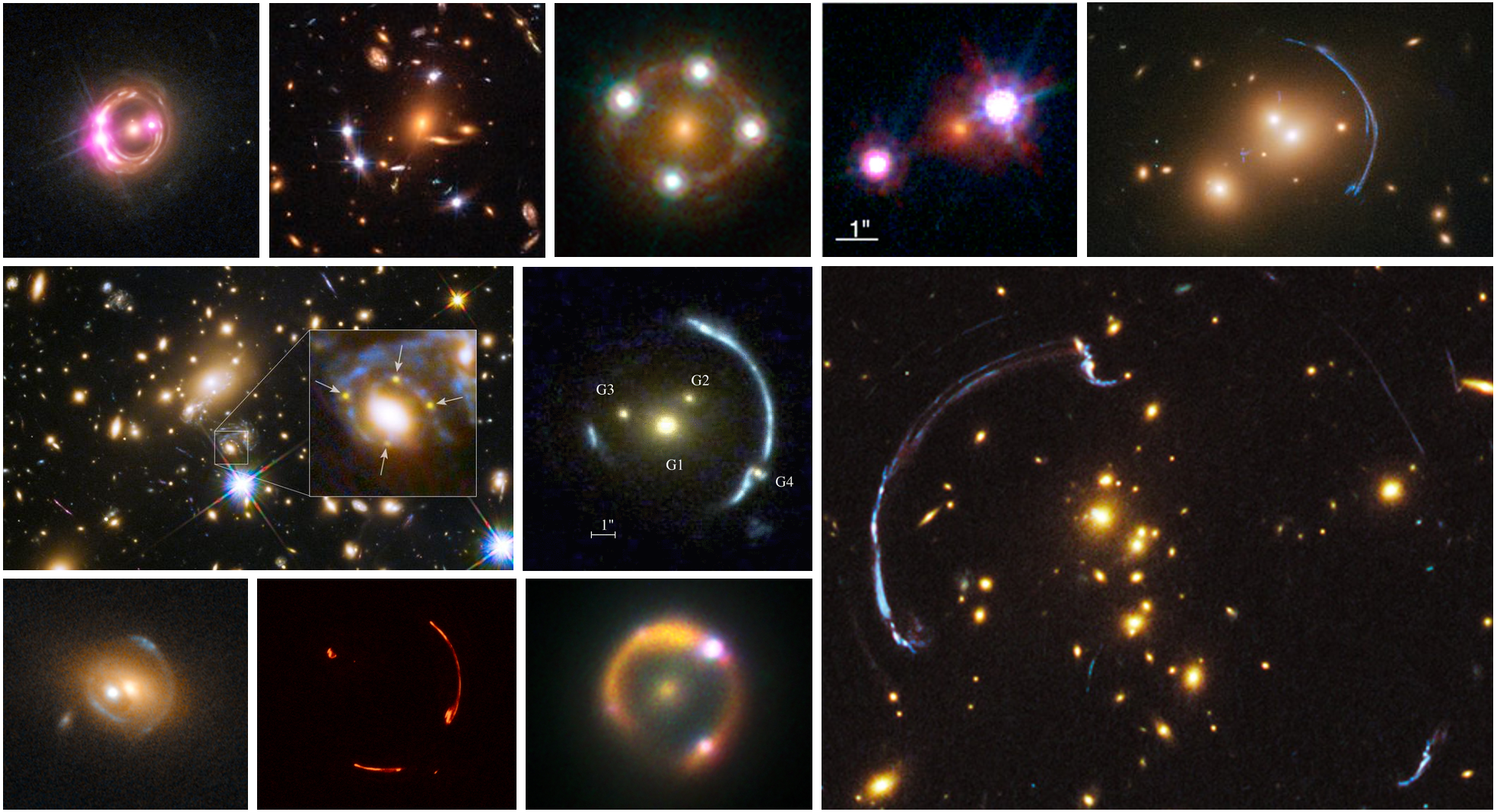}
    \caption{Various strong lens systems with quasar, galaxy, and supernova sources. Images are composed from multi-band \textit{HST} imaging, a mix of \textit{HST} and ground-based adaptive optics, or Very Long Baseline Interferometry (VLBI). Sources without optical, infrared, sub-mm, or radio imaging counterparts should not be forgotten, however no definite localised examples have yet been discovered. Credit: NASA/ESA/NRAO/AUI/NSF.}
    \label{fig:lens_collage}
\end{figure*}

Gravitational lenses span a wide range of characteristics, with various brightnesses, image configurations, separations, lens types, source types, and environments. Figure \ref{fig:lens_collage} shows a gallery of gravitational lenses to demonstrate this variety of possibilities. Before describing the lens search techniques (Section \ref{sec4:seclection_methods}), it is first prudent to describe the physical characteristics of the possible lens systems themselves. For simplicity, we split the descriptions between the deflector (this Section) and source populations (Section \ref{sec3:classes_source}). We limit our discussion to \textit{macro-lenses}, in which the multiple images of the source are separated by the order of arcseconds, and can be readily resolved by current observatories. Such separations are only possible for deflectors and sources at cosmological distances, and with masses above $10^8$~M$_{\odot}$ so the relevant lensing objects are isolated galaxies (early- and late-type), galaxy groups, and galaxy clusters.

For lens-selected searches, an initial catalogue of systems is based on finding examples of the expected lens type. For example, high-redshift galaxies lensed by galaxy clusters might be found through an all-sky selection of galaxy clusters. Below we describe the physical properties of these deflector classes to better understand the lens search techniques and their possible biases.

%Lens-selected search methods are tuned to go after specific lens types, \chiara{in particular single galaxies or groups/clusters of galaxies. Here in this section, we highlight the two types, focusing on the key observables helping us to find lenses. In the next section we focus instead on the sources type, while in Section~\ref{sec4:seclection_methods} we describe the different selection methods used for one or the other class.} 

%, hence we describe lenses here. What are the different methods to find them? What are the key observables helping us to find lenses?

%Somewhere we need a sentence that mentions;

%The superposition and relative contributions of the fluxes from the foreground lens and the background source each at different redshifts and with different rest frame spectral energy distribution ....... 
%Mention spatial resolution. Maybe show a HST and ground based image of a lensed quasars; some cluster arcs. Need a rouges gallery showing ground based discovery data images and higher resolution HST image \shu{(what about Figure 1?)}; also AO; radio images too.

%%%%
\subsection{Galaxies}%Finding galaxy-scale lenses}
%\chiara{LET'S HAVE A LOOK AT CHAPTER2 - THEIR INTRO }
Galaxy families are often split by morphology; we can broadly categorise them into ellipticals (early-type), and spirals (late-type) via the Hubble sequence. Ellipticals are featureless quiescent galaxies with old stellar populations and no star formation, representing the final product of hierarchical galaxy formation. Spirals, on the other hand, show prominent star formation and coherent rotation, in contrast to pressure-supported ellipticals. Due to their larger numbers at high masses, ellipticals dominate over spirals as lensing galaxies (approximately 7 to 1), with the latter only contributing at image separations smaller than 2 arcseconds \citep{turner1984, kochanek1996}. For more details on lensing by galaxies, we refer the reader to the Galaxy Lensing Chapter.

\sloppy
\subsubsection{Early-type lenses}
\begin{figure*}[htbp]
    \centering
    \includegraphics[width=0.98\textwidth]{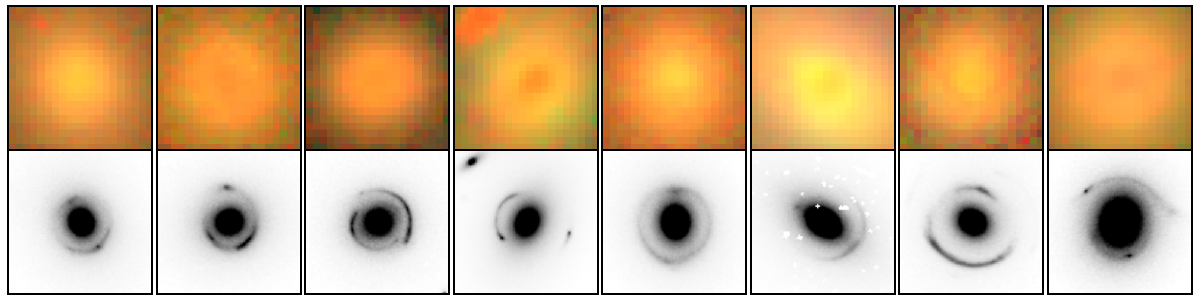}
    \caption{Ground-based \textit{grz} images from the Legacy Survey (top) and \textit{HST}-ACS WFC F814W (bottom) images of eight galaxy-galaxy strong lenses from the SLACS sample \citep{bolton2008}. The images are ordered by Einstein radius from left to right. Each cutout is 5$^{\prime \prime}$ across.}
    \label{fig:slacs_ground_vs_HST}
\end{figure*}

Elliptical galaxies, often found within groups and clusters, have ellipsoidal isophotes, are composed of old stellar populations, and lack significant star formation, dust, or gas. They are accordingly red, with the most massive examples termed Luminous Red Galaxies (LRGs). Due to both selection effects of brighter galaxies, and their increased lensing efficiency, massive LRGs constitute the majority of known high-confidence galaxy-scale lenses, and dominate the system's flux (and therefore colours) in ground-based imaging surveys, as shown in Figure \ref{fig:slacs_ground_vs_HST}. Ellipticals' spectra contain clear absorption lines from old stars (Ca\textsc{ii} H and K, Mg-b, and Na-D), and are void of emission lines. A typical feature used to measure redshifts of ellipticals is the characteristic 4000\AA\ break immediately redward of the Ca\textsc{ii} H and K lines. The light profiles of LRGs are often well-described by a single Sersic profile (with Sersic index $n\approx4$), however there is evidence for a dichotomy of formation pathways for massive ellipticals, with one coreless population with extra central light and lower Sersic index fits \citep{kormendy2009}. 

%A Sersic profile alone has been shown to be insufficient for the baryonic component of the light, further.

As ellipticals dominate the galaxy-scale lensing population, several lensing studies have constrained their mass distributions, revealing standard NFW halos, heavy initial mass functions \citep[e.g.][]{Sonnenfeld2015,shajib2021}, and an alignment between total mass and light \citep[e.g.][]{keeton1997, shajib2019, schmidt2022}. Their total mass profiles are approximately isothermal, despite the relative non-isothermal contributions from baryons and dark matter \citep[the so-called `bulge-halo conspiracy', e.g., ][]{treu2004, auger2010, dutton2014}, however detailed individual lens studies prefer multiple baryonic components or departure from purely elliptical masses \citep[e.g.,][]{nightingale2019, powell2022}. The velocity-dispersion function (VDF) has been found to be consistent with no redshift evolution below $z\sim0.5$ \citep{shu2012}, however, the evolution beyond that is unclear, as large samples of spectroscopic measurements are currently limited to below $z\approx1$. Complete gravitational lens samples, through their image separation statistics, are a sensitive probe of the elliptical VDF and its evolution \citep{oguri2012}.

\subsubsection{Late-type lenses}
The mass distribution of spiral galaxies can be decomposed into their disk, bulge, and halo. The relative mass contributions of each component is unclear from kinematic data alone due to the so-called `disc-halo degeneracy' \citep{vanalbada1986}. Gravitational lensing offers a unique method to break this degeneracy given the additional mass constraints from lensing, and constrain both the initial mass function and halo properties \citep{barnabe2012}. 

\begin{figure*}[htbp]
    \centering
    \includegraphics[width=0.98\textwidth]{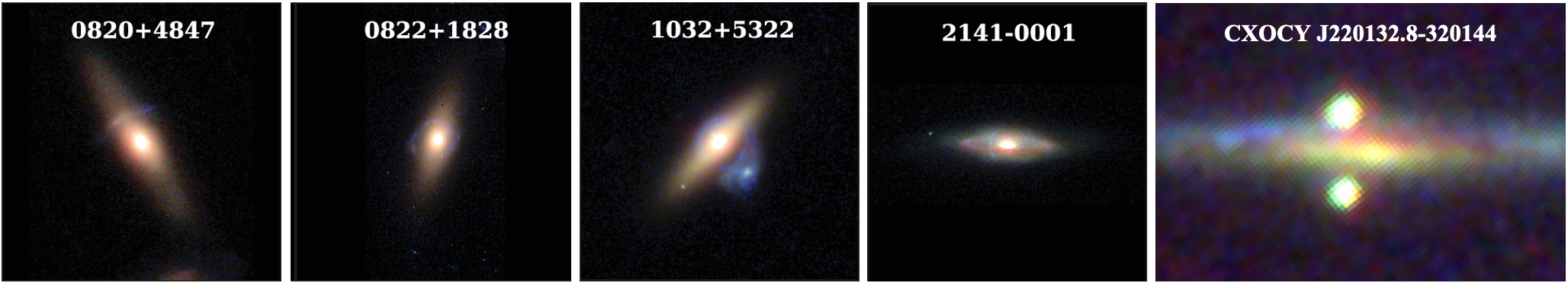}
    \caption{Examples of lenses with late-type galaxy deflectors; the first four are from the SWELLS sample \citep{Treu2011}, and the last one is a serendipitously discovered lensed quasar \citep{castander2006}.}
    \label{fig:latetype}
\end{figure*}

The efficiency (or cross-section) for lensing by these systems is strongly dependent on the inclination, as this significantly changes the projected mass density --  a key quantity for lensing probabilities and configurations. \citet{keeton1998} show that the bulge component is necessary for strong lensing to occur in face-on systems, which would otherwise not surpass the surface mass density required for multiple imaging, and quad systems are expected almost exclusively from edge-on systems. A handful of serendipitous discoveries and a concerted search within SDSS spectra -- SWELLS \citep[Sloan WFC Edge-on Late-type Lens Survey,][see also Section \ref{fibrespectra}]{Treu2011} -- now provide $\sim$ 20 such examples. Figure \ref{fig:latetype} shows a selection of these lenses. The presence of dust is common in these galaxies, and the extinction of both point sources and extended sources must be accounted for during searches and modelling.

%Galaxies, and in particular massive early-type galaxies, are ideal for strong lensing of distant sources as they are abundant and their mass distribution is sufficiently concentrated. %CS:citation?  
%Sometimes edge-on and, in rare circumstances also face-on spiral galaxies have been found to produce strong lenses {\bf ADD REF}. Both imaging and spectroscopy based techniques have been successful in finding galaxy-scale lenses as these are relatively simple systems, as we describe in this chapter. These techniques rely on features specific to the deflector as well as the characteristics of the lensed sources {\bf add ref to subsequent sections}. 
%Lensing \chiara{provides a one-shot, only gravity-dependent, very precise estimate of the total mass of the deflector. When combined with other probes, such as stellar kinematics,  dynamics and/or stellar population analysis, lensing can help in constraining the luminous versus dark matter distribution {\bf ADD REF:Leon and Tommaso}, the low-mass end of the initial mass function (IMF) {\bf ADD REF: Chiara and Alessandro Sommerfeld}, as well as }%can help constrain the mass distribution, stellar initial mass function (IMF), dark matter fractions and other statistical properties such as the ellipticity and so on {\bf update these later}.

%%%%
\subsection{Groups and clusters} \label{clusters}
Galaxy groups are loosely defined as comprising several to $\sim$ 50 galaxies within $\sim1$ Mpc, and are thought to be embedded within their dark matter halos. Galaxy clusters contain anywhere upwards from $\sim$50 members, and can span up to 10 Mpc. %Groups tend to have masses $\sim10^{13-14}$~M$_{\odot}$ and clusters have masses over $\sim10^{14}$~M$_{\odot}$.
Equivalently, groups have masses $\sim10^{13-14}$~M$_{\odot}$, and clusters have masses over $\sim10^{14}$~M$_{\odot}$.
Such clusters can lens many dozens of high-redshift sources, making them efficient tools for studying high-redshift galaxies individually or in a statistical sense. While these extra sources provide more constraints \citep[e.g., 30 sources with a total of 90 multiple images in Abell 2744,][]{bergamini2023}, the mass distributions are naturally also more complex than isolated galaxies, and even state-of-the-art models are limited to reproducing image positions of multiply sourced systems to within a few tenths of an arcsecond in the image plane \citep[e.g.,][]{acebron2022}. 
As a result, the lensing configurations are more often non-standard \citep[e.g.,][]{orban2009}, often leading to larger source magnifications than isolated galaxies \citep[e.g.,][]{robertson2020}, and the lensed images have wider image separations than produced by galaxy-scale lenses \citep[e.g.,][]{more2012}. Complex configurations are particularly common in disturbed clusters with ongoing mergers, reflective of the turbulent mass distributions. At cluster scales, the lack of baryonic cooling beyond a certain mass scale implies fundamentally different mass profiles relative to isolated galaxies. X-ray surface brightness and strong and weak lensing analyses show that the total density profiles of relaxed clusters are well-described by Navarro-Frenk-White (NFW) profiles, namely an inner density slope of $\alpha=1$ \citep[$\rho(r)\propto r^{-\alpha}$;][]{schmidt2007, newman2013}. This leads to less efficient lensing in clusters, and explains well the paucity of observed cluster-scale lensed quasars \citep{keeton1998b, kochanek2001}. The associated shallower mass profiles lead to a predicted surplus of cusp geometries, in which only three bright images are apparent, and coupled with triaxiality, this can reduce the lensing efficiency by up to a factor of 4 \citep{oguri2004}.

We briefly discuss the discovery methods for clusters, since this is often a significant step towards finding new giant arcs. The hot intracluster gas emits X-ray emission, which has been used to discover clusters in all-sky X-ray surveys, most notably the MAssive Cluster Survey \citep[MACS,][]{ebeling2001}. In optical imaging surveys, clusters are found as overdensities in the 4-dimensional space of position, colour, and magnitude, for example in the Red-Sequence Cluster Sequence \citep[RCS,][]{gladders2005} and the SDSS Giant Arcs Survey \citep[SGAS,][]{sharon2020}. The available datasets for these catalogues are key to the discovery of multiply imaged sources. The deepest datasets -- such as those of the Frontier Fields \citep{lotz2017} and the Reionization Lensing Cluster Survey \citep[RELICS,][]{coe2019} -- reveal sources as intrinsically faint as 33 mag.

%{\bf add ref to SARCS More et al. and CASSOWARY samples}.
%At cluster scales, the lack of baryonic cooling beyond a certain mass scale can be important for explaining the paucity of observed cluster-scale lensed quasars \citep{keeton1998b, kochanek2001}. The associated shallower mass profiles lead to a predicted surplus of cusp geometries, in which only three bright images are apparent, and coupled with triaxiality, this can reduce the lensing efficiency by up to a factor of 4 \citep{oguri2004}. 
%In the more massive clusters, the presence of the central brightest cluster galaxy, the triaxiality of the halos and increased number of satellite members (or substructure) contribute to the increased lensing cross-section producing numerous background lensed galaxies per lensing cluster. 
For a full description of lensing by clusters, we point the reader to the Cluster Lensing Chapter.

%Finding groups- and cluster-scale lenses}
%Galaxy groups and clusters are gravitationally-bound massive structures comprising of a few tens to hundreds of individual galaxies embedded in their respective dark matter halos. While there is no clear distinction between groups and clusters, 

%Since the number density of groups and clusters is smaller than galaxies, somewhat different approach is used for finding lens systems at these massive scales. Several algorithms have been developed for finding galaxy groups or clusters from imaging surveys with multi-band data {\bf ADD REF}. These algorithms make use of colors or over-densities of galaxies to search for these systems. Subsequent to this step, different methods have been proven to be efficient. For instance, images of groups and clusters are visually inspected to look for lensing signatures. Alternatively, as these systems tend to show giant arcs, some of the search algorithms focus on searching for these prominent and elongated arcs as a means for finding groups- or cluster-scale lenses {\bf ADD REF}.

%Strong lensing of groups and clusters is used to study dark matter and total mass distribution {\bf ADD REF}, to probe cosmology via the abundance of clusters  {\bf ADD REF}, to probe dark energy via lensing tomography  {\bf ADD REF, ADD MORE applications later}. 

%%%%
%\subsection{Clusters of Galaxies}

%%%%%%%%
\section{Lensed source classes}
\label{sec3:classes_source}

%%%%
We briefly review the various sources of gravitational lenses, highlighting the physical characteristics that are important for their discovery and application to lens searching methods.

%\chiara{}
%%%%
\subsection{Galaxies}
The deflector classes from Section \ref{sec2:classes_deflectors} can naturally be the sources of lower redshift deflectors. However, ETG sources are intrinsically rare, with only 15 known examples of `Early-type Early-type Lenses' \citep[EELs, systems in which both the lens and source are ETGs,][]{Oldham2017}. While the mass function was more important when discussing the deflector population, the luminosity functions and source shapes are the key metrics for the source populations. We cover two general classes of common source galaxies: unobscured, blue star-forming galaxies and dust-obscured star-forming galaxies (or sub-mm galaxies). 

\subsubsection{Blue star-forming galaxies}
In the hierarchical picture of galaxy formation, galaxies are built up through mergers and accretion of smaller galaxies, accompanied by the onset of star formation. This star formation is seen to peak around `Cosmic Noon' ($z\approx2$), in which the merging galaxies are predominantly low-mass, late-type galaxies -- irregulars and spirals. Such high-surface-brightness, compact galaxies dominate the galaxy population, and particularly the observed-frame optical luminosity function, and thus represent one of the most common lensed sources of gravitational lenses. 

The spectrum of a star-forming galaxy composes a blue continuum with absorption lines from gas proximate to the star-forming regions, and any associated emission/nebular lines, such as [\textsc{Oii}] 3727 and [\textsc{Oiii}] 5007. These features are particularly useful for identification in spectra (see Section \ref{fibrespectra}). At optical wavelengths, requiring the detection of multiple common emission lines typically limits the redshifts of such sources to $z\lesssim1$. Beyond this, single emission lines can only suggest a redshift, however at $z\gtrsim2$, Ly$\alpha$ enters the optical, and may be distinguished at sufficient signal-to-noise as its profile often has a sharp edge towards the blue, and a tail towards the red \citep{Shu2016}.

\begin{figure}
    \centering
    \includegraphics[width=\textwidth]{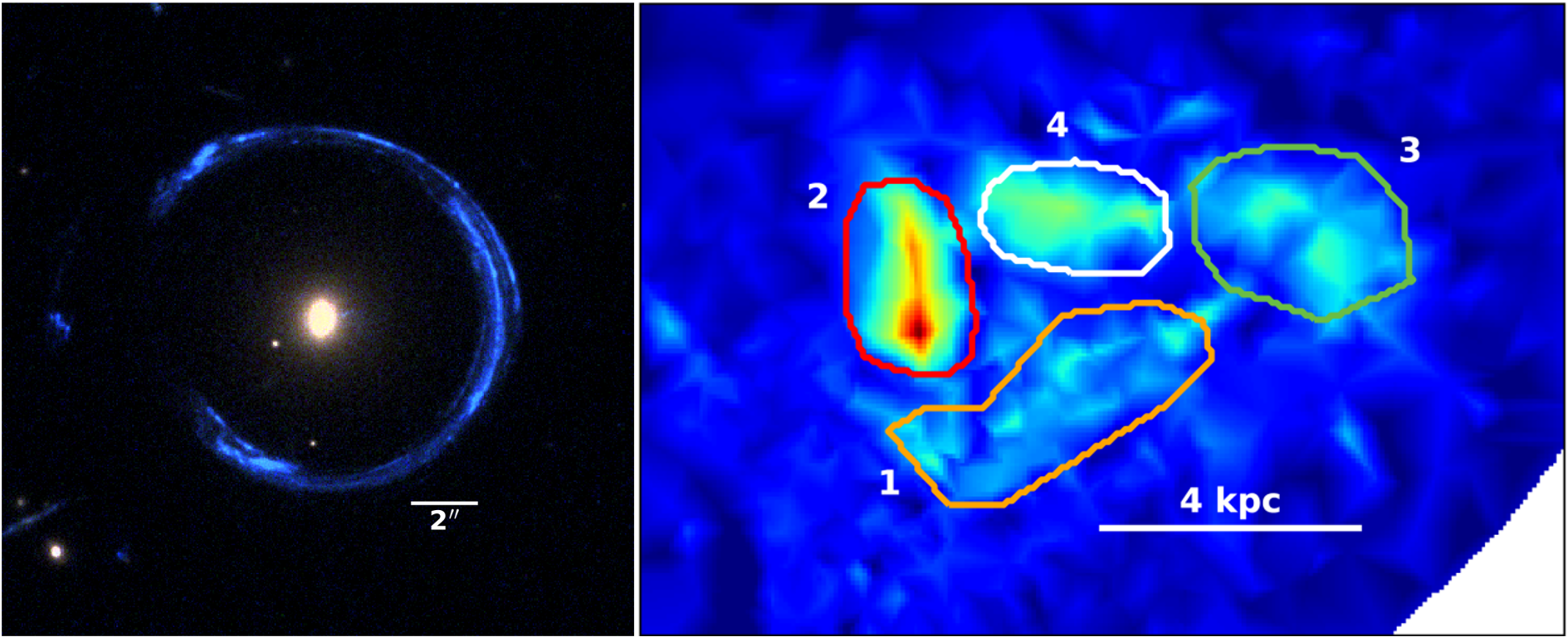}
    \caption{\textit{Left}: \textit{HST} F475W/F606W/F814W colour image of the Cosmic Horseshoe; \textit{right}: reconstructed source ($z=2.38$) from the F160W observation, showing four distinct star-forming regions. Reproduced from \citet{james2018}.}
    \label{fig:horseshoe}
\end{figure}

The shapes of these galaxies compose multiple star-forming regions perhaps due to recent mergers, with several compact and diffuse components \citep{ritondale2019}. For example, Figure \ref{fig:horseshoe} shows the well-known lensed star-forming galaxy, the Cosmic Horseshoe \citep{belokurov2007}, with the reconstructed source showing several typical clumps.

%Late-type galaxies -- spirals and irregulars -- are typically more abundant and show more star formation their early-type galaxy counterparts -- ellipticals and lenticulars. This star formation is onset by

\subsubsection{Dusty star-forming galaxies}
Dusty star-forming galaxies (DSFGs), also known as submillimeter galaxies, are high redshift (median $z\sim4$) galaxies with remarkably bright blackbody emission in the rest-frame far infrared, which is redshifted into the millimeter and submillimeter bands. These galaxies are some of the most active star-forming galaxies in the Universe, with star-formation rates exceeding thousands of solar masses per year \citep[e.g.,][]{casey2014}. Their star-forming regions are obscured by dust, which, after being heated by UV radiation, are responsible for the bright far-infrared emission. %Given both their high-redshift and remarkable star-formation rates, these systems represent powerful tools for testing star- and galaxy-formation models in the early Universe, however, detailed observations of DSFGs have been challenging both due to detection sensitivity and the difficulty of spatially resolving their structure. Strong gravitational lensing has been a powerful tool to overcome these challenges. %By boosting the total observed flux of these sources, strong lensing has allowed the detection of faint signals from these sources. Additionally, by spatially magnifying the images of these galaxies, strong lensing has allowed resolving detailed structures as small as XX parsec (e.g., cite) in these high redshift sources. 
In addition to copious amounts of dust, these galaxies contain vast reservoirs of molecular gas, which are commonly observed through their emission lines \citep[e.g.,][]{Weiss2013}. In contrast, these galaxies are notably difficult to detect in rest-frame optical or UV-bands, as such emission is absorbed and scattered by dust. 

One of the unique properties of DSFGs is the fact that their (unlensed) luminosity function has a sharp cut-off at the bright end \citep[e.g.,][]{negrello2010}. This simply means that (without lensing) there are large numbers of lower-flux galaxies and almost no extremely bright sources. Therefore if an unresolved extremely bright DSFG is observed, it is highly likely to be strongly lensed due to the associated magnification (see Section \ref{magnification_bias}). Since most lensing galaxies are faint in the submillimeter, such lensed DSFGs do not suffer from lens-source deblending problems that are common at optical and infrared wavelengths.

%In the past decade, a number of wide-area mm and submm surveys observed large areas of the sky, identifying a population of unresolved sources with submm colors consistent with blackbody emission by dust.  Statistical lensing studies indicated that the distribution of the observed brightness of these sources was consistent with the predictions of statistical lens models. High-resolution follow-up observations (primarily by ALMA) then revealed that these sources were indeed primarily composed of strongly lensed DSFG galaxies. 

\subsection{Quasars}

%%% Added by LD on 2022-01-11 %%%

Supermassive black holes are thought to exist in the central potential of all massive galaxies, and can have luminosities of up to 10$^{15}L_{\odot}$, due to the thermal emission associated with friction in the accretion disk. While in this actively accreting state, they are known as Active Galactic Nuclei (AGN), and the brightest high-redshift (typically $z>1$) examples are quasars, which offer a common source population for gravitational lenses, given their brightness and luminosity function peak at high-redshift.

\begin{figure}
    \centering
    % LD: I took this figure from my PhD thesis, Reference: http://isdc.unige.ch/3c273/
    \includegraphics[trim={0 0 0 4mm},clip,width=\textwidth]{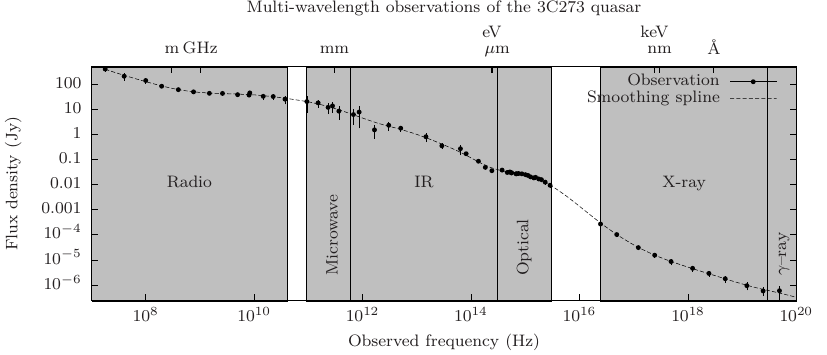}
    \caption{Multi-wavelength observations of the 3C273 quasar at $z = 0.158$ \citep[data taken from][]{turler1999}.}
    \label{fig:3C273_wall}
\end{figure}

Quasars emit over the whole electromagnetic spectrum, as shown in Figure \ref{fig:3C273_wall} in the case of 3C273, the apparently brightest quasar on the sky. We point the reader towards \cite{2012agn..book.....B} and \cite{padovani2017} for explanations of the physical processes that cause the multi-wavelength emission of quasars. In brief, the accretion disk contributes a continuum power-law in the rest-frame UV to infrared, and surrounding broad and narrow line regions of ionised gas add broad and narrow lines attributed to specific ions. Strong radio emission is found in about 10\% of quasars, and is attributed to the presence of powerful jets that can extend out to many times the size of the host galaxy. A variety of observed quasar `types' are possible depending on viewing angle: when viewed face-on, sources have prominent blue continua and broad emission lines (type-I); edge-on viewing leads to an obscured accretion disk and narrow lines (type-II). This continuum of viewing angles leads to a variety of observed quasar properties, and is further divided by levels of dust in the host galaxy, the presence of radio jets, and absorption due to outflows.

A typical scale for an accretion disk is several light days, thus quasars are unresolved at cosmological distances, and appear as point sources often outshining their host galaxies. To select these systems without contamination from stars, several wavelength-specific selection techniques have been used: UV-excess selection \citep{2009ApJS..180...67R}, unique infrared colours \citep{stern2012}, and radio and X-ray detections. Quasars are also seen to be significantly variable at all wavelengths on timescales above a week. This can be used as a selection method, with the variability amplitude being found to positively correlate with bluer rest-frame wavelengths and smaller black hole masses \citep{macleod2010}. We note that UV-excess selections are not reliable for higher quasar redshifts, as the Lyman alpha dropout appears in the UV filters, leading to colours similar to stars. Equivalently for infrared colour selections, the local minimum at $1 \mu$m is redshifted into redder filters, increasing stellar contamination.

% LD: I've put the itemize list below a long time ago but it is not relevant anymore such that I commented it out
% \begin{itemize}
%    \item Make a smooth transition from lensed galaxies to lensed quasars (i.e. core dominated)
%    \item We are searching for multiple point-like, variable sources with a steady astrometry
%    \item How can we do it ({\bf to be completed})
%    \begin{itemize}
%        \item Using colour around known/candidate QSOs (e.g. SQLS)
%        \item Combining ground-based and space observations (e.g. Gaia multiplet around DES quasar candidates)
%        \item Variability-based (e.g. searching for low entropy light curves)
%        \item `Morphological' approach (e.g. searching for image configuration that are compatible with a lensed system)
%    \end{itemize}
%    \item Would a plot of redshift vs. angular distance (or any other cosmological distances) be useful here?
%\end{itemize}
%%%%%%%%%%%%%%%%%%%%%%%%%%%%%%%%%

%\noindent General features that should be described in this subsection:
%\begin{itemize}
%    \item spectral characteristics/different types (type I, II, BALs, reddened, etc.)
%    \item description across wavelengths: (how many have radio/X-ray detections, known correlations(?), redshift dropouts, photometric catalogues)
%    \item variability features
%    \item luminosity function across redshift perhaps?
%\end{itemize}

%%%%
\subsection{Transients}
Transients cover all variable astrophysical phenomena which eventually fade away on human timescales. While supernovae have been known for millennia, several new classes of transient have only been discovered in recent years, and our current understanding suggests that they should exist at high-redshift offering sources for gravitational lenses. The discovery of the first definite lensed examples is only a matter of time as observatories open up new volumes and resolving capabilities. For complete reviews of lensed transients, we direct the reader to \citet{oguri_transient_review} and \citet{liao2022}.

\subsubsection{Supernovae}
Supernovae are intrinsically rare events, constituting the final moments in the evolution of only the most massive stars or specific binary star systems. The former are known as Type Ia supernovae and the latter are core-collapse supernovae. A wide array of supernova subtypes exist based on the presence, or lack thereof, of various narrow absorption lines and exact lightcurve shape. All supernova lightcurves are smoothly varying with a maximum brightness at 15-25 days after explosion, followed by a slow fade, and typical supernovae rates are roughly once per galaxy per century. \citet{refsdal1964} first suggested supernovae as sources in gravitational lens systems, highlighting that their time variations allowed precise determination of a time delay, and in turn, a measurement of the Hubble constant. 

For detection of such transients, the Zwicky Transient Facility has played a pivotal role by surveying the entire sky every two nights down to a magnitude of $r\sim20$ thanks to its mosaic camera, capable of imaging 47 square degrees of sky at once. From 2016 to 2023, over 100,000$^{1}$\blfootnote{$^{1}$ Based on searches of the Transient Name Server: https://www.wis-tns.org/stats-maps} transients consistent with supernovae have been reported by ZTF and various cadenced imaging surveys. However only 10\% of these have spectra, often limited to those brighter than $\sim$18.5 mag.

The standard-candle nature of Type Ia supernovae makes them valuable not only for cosmological probes \citep[see, e.g., ][for a review]{2011ARNPS..61..251G}, but also as a method for detecting an associated magnification from strong lensing. Only a spectroscopic redshift, classification, and peak magnitude are needed (see Section \ref{magnification_bias}). We refer to the reader to the Lensed Supernova Chapter for a full description.

\subsubsection{Fast radio bursts}
Fast Radio Bursts (FRBs)~ are a new class of radio transients at cosmological distances whose durations range from nano- to milli-seconds. Around 100 FRBs have been discovered, and their population-inferred redshift peaks at $z = 0.4$ \citep{shin2022inferring}. The individual redshifts of most bursts are unknown since they are unlocalised, though a handful have been localised to $z>0.5$ galaxies \citep[e.g.,][]{ryder2022probing} . The physical origin of FRBs is unknown, however they are divided into repeating and non-repeating sources. We point readers to \citet{petroff2019fast} and \citet{cordes2019fast} for detailed reviews.

\subsubsection{Gamma ray bursts} 
Gamma Ray Bursts (GRBs) are brief highly energetic explosions that appear at gamma ray frequencies and display afterglows at all wavelengths. They are divided into long (70\%) and short (30\%) GRBs, separated by durations around $\sim$2 seconds \citep{kouveliotou1993}. Long GRBs are associated with supernovae, and short GRBs with neutron star mergers.

Since GRBs are particularly bright and occur at cosmological redshifts, they are expected to be gravitationally lensed \citep{oguri2019}. Of order $10^{4}$ GRBs have been observed, one third of those with the Fermi Gamma-ray Space Telescope. Were the redshift distribution for GRBs the same as that for quasars, one would expect several of these to have been gravitationally lensed. The search techniques and potential discoveries are presented in Section \ref{transient_temporal}.

\subsubsection{Gravitational waves} 
Since the first detection of gravitational waves (GW) from the merger of binary black holes in 2015 by the Laser Interferometer Gravitational-wave Observatory (LIGO), nearly 100 GW events from binary systems have been detected. These binaries comprise either a pair of black holes (up to 100$M_{\odot}$), neutron stars ($\sim 2$M$_{\odot}$), or a neutron star and a black hole. The current ground-based GW observatories, LIGO--Virgo--Kagra, are sensitive to GW signals in the frequency range $10-10^3$~Hz and the angular accuracy of the sky localisation contours spans an area of $\mathcal{O}(100)$~deg$^2$. Supermassive black hole mergers (up to $10^{10} M_{\odot}$) will emit in the nano-Hz frequency range, in which Pulsar Timing Arrays (PTAs) are sensitive \citep{nanograv2023, xu2023}. Gravitational lensing of GWs is speculated to be discovered within the next few years with much theory already laid down a few decades ago \citep[e.g.][]{DeguchiWatson1986,Takahashi1998}.

%Gravitational waves -- perturbation in the space-time continuum -- are another important prediction of Einstein's Theory of Relativity other than gravitational lensing. 

  %Among the detected sources of compact binaries, we have discovered neutron stars ($\sim 2$M$_{\odot}$ onwards) to stellar-mass black holes $(\mathcal{O}(100)$M$_{\odot}$). 

Binary black holes (BBHs) are expected to be the first lensed GW detections owing to their cosmological distances. The exquisite precision on the time delay between multiply lensed BBH will make them an attractive probe of cosmological parameters such as the Hubble constant \citep[e.g.][]{Liao2017}. This is feasible via their distance measurements when combined with the complementary electromagnetic (EM) data constraints, for instance, optical redshifts, accurate lensed image positions and lensing configuration. If the binary has a direct EM counterpart as is expected in the case of neutron stars (e.g. GW170817, the first BNS), lensing can shed light on the physical processes right before the merger through an early warning study \citep[]{Magare2023}. 

\begin{sidewaystable} 
\vspace{12cm}
\renewcommand{\arraystretch}{2}
%\begin{table}[h!]
\caption{Summary of discovery methods and numbers of known lens systems. Text in italics implies that the method is a prospective idea for a search.}
\centering
\fontsize{10pt}{10pt}\selectfont
\begin{tabular}{| p{15mm} | p{35mm} | p{35mm} | p{35mm} | p{30mm} | c | } 
 \hline
\centering Source & \centering{Magnification Selection} & \centering{Multiple Imaging Selection} & \centering{Multiple Redshift Selection} & \centering{Time-delay Selection} & Number \\
 \hline
\centering Galaxy & very pure for DSFG sources given steep bright-end of luminosity function & visual inspection, citizen science, ring-finders, neural networks & narrow emission lines in galaxy fibre searches / IFU searches & \centering{N/A} & $\sim$1500 \\ 
\hline
\centering Quasar &  UV-X-ray luminosity relation / proximity zone size from Ly$\alpha$ transmission / Eddington ratio & catalogue-only from \textit{Gaia} / deblended SED similarity / extended variability & rare, serendipitous broad emission lines in galaxy fibre spectra & rare, one discovery from OGLE lightcurves & $\sim$300  \\
\hline
\centering Supernova & spectroscopic supernovae that are too bright / \textit{inconsistent with photometric redshift of nearby elliptical} & multiple detections in difference images of \textit{HST} monitoring of clusters & \centering{N/A} & \textit{monitor known supernovae for re-brightening} & 5 \\
\hline
\centering GW \\ FRB \\ GRB & \centering{N/A} & \textit{resolved images of EM counterparts} & \centering{N/A} & \textit{require similar waveform and sky localisation of time-separated detections} & 0 \\ 
 \hline
\end{tabular}
\label{table:discovery_summary}
%\end{table}
\end{sidewaystable}

%%%%
\section{Selection methods}
\label{sec4:seclection_methods}
%\chiara{The power of gravitational lenses as probes of astrophysics and cosmology has fuelled an array of searches across a variety of datasets and wavelengths. To find lenses, one has to separate the relative (superimposed) contributions of the fluxes from the foreground lens and the background source, each at different redshifts and with different rest frame spectral energy distribution. One} 
One single search method for all lenses is impossible, given the wide variety of lens types, source types, image configurations, colours, available surveys, etc. Therefore lens searches are tailored to target specific lenses in specific datasets. Search techniques can be fully described by: (i) the type of lens being searched for, (ii) the data being used, and (iii) the algorithm that prioritises or selects lens candidates. These algorithms often target one of four unique features of gravitational lenses: source magnification, multiple well-separated images, multiple redshifts, or time delays. We divide this section by these lensing attributes. For example, a discovery method that searches for multiple blue arcs around a red galaxy requires only the second feature, multiple imaging, and is independent of the other three. We remind the reader, however, that modern selection techniques can rely on a combination of these methods. We describe serendipitous discoveries in their own subsection, and citizen science searches within the multiple imaging section. Machine learning techniques are given their own section though principally rely on the resolved-images feature of lensing. A summary of the various discovery methods, divided by source type, are listed in Table \ref{table:discovery_summary} with approximate numbers of known systems.

\subsection{Serendipitous discoveries}
Nearly all of the first dozen lenses were serendipitous discoveries, and even now lenses continue to be found fortuitously, and merit single publications. We do not aim to describe all these discoveries but highlight some of the most interesting and important examples. 

The first known lens, Q0956+561, was part of a follow-up programme for quasars (at that time defined as radio sources) from the Jodrell Bank radio survey, to identify optical counterparts, and determine redshifts \citep{walsh1979}. They obtained spectra of two blue point sources that might be associated to the radio emission, separated from one another by 6 arcseconds. The spectra were close to identical (at $z=1.405$) and soon after the lensing galaxy was discovered \citep{young1980}. Within a year of this first lens discovery, a known quasar, PG1115+080, was being followed up for high resolution spectroscopy, when two further stellar objects were observed on the TV guiding monitor within $\sim$3 arcseconds. \citet{weymannn1980} proposed this as a lensed triple (at $z=1.722$), and later observations showed that the brightest image A was in fact a close pair of images \citep{hege1981}. A few more years would pass before a third gravitational lens was discovered, also by serendipity. Q2237+0305, also known as the Einstein Cross, was discovered by \citet{huchra1985}, during a spectroscopic survey of bright galaxies. The spectrum showed a quasar at $z=1.7$ at the nucleus of a $z=0.039$ spiral galaxy, and later imaging revealed the four individual quasar images \citep{yee1988}. Lenses that are outliers in terms of their extreme properties, such as the very low redshift lens of Q2237+0305, are, perhaps unsurprisingly, often found serendipitously. For example, one of the brightest known lensed quasars, APM08279+5255, at $z=3.87$, was a serendipitous discovery within a survey of distant cool carbon stars \citep{irwin1998}, which share similar infrared colours to high-redshift quasars. Also, the lensed quasar with the lowest source redshift, RXJ1131$-$1231, was discovered serendipitously while measuring polarisations of unlensed quasars \citep{sluse2003}. Even in the last few years, bright quadruply lensed quasars are discovered as a byproduct of visual inspection -- the two diamonds of \citet{lucey2018} were identified when removing contaminants for the target catalogue of the Taipan Galaxy Survey.

%In the original quasar catalogue, a star was misidentified as a quasar likely because it was a point source while the actual lensed quasar was not, and therefore not matching the original definition of a quasar requiring it to be a point-source. 

Many lensed galaxies have also been discovered unexpectedly, including many of the earliest examples. The giant arc in the galaxy cluster Abell~370 was noticed on the CCD images during a spectroscopic campaign of the galaxy members \citep{soucail1987}. Originally it was thought to be due to tidal interactions or star formation, but soon after, spectroscopy placed the source at $z=0.59$, i.e. lensed by the $z=0.374$ cluster. Galaxy-scale complete Einstein rings are striking and thus relatively easy to spot even within a wide field of data, and are thus regular serendipitous discoveries in the literature \citep[e.g.,][]{tanaka2016, bettinelli2016}, and similarly for bright galaxy-scale arcs \citep[e.g.,][]{allam2007}.

\subsection{Magnification selection}
A general property of strong gravitational lensing is the \textit{magnification theorem}: at least one image is as bright as the source \citep{schneider84}. Therefore a magnification $>1$ is expected in any strong lens system, implying that the brightest sources will contain an enhanced fraction of lenses. Some systems will show anomalously bright sources that would disobey limits imposed by physics if not lensed. We discuss these two consequences as lens search techniques below.
%CS: not true for arcs! A common feature of all strong gravitational lenses is multiple images of the background source. One can therefore look for the unique configuration of lensed images around the lensing galaxy in imaging data. 

%Include discussion on HE and Liege targeting surveys
%C.L.: I don't think this is the right way to introduce this section? The blue arcs stuff isn't relevant to magnification bias targeting (at least I don't know of it in practice?) and isn't necessarily true either. It would be better to include in multiple image search. I can rewrite this.
%\chiara{A common feature of all strong gravitational lenses is the presence of the background source around the deflector, which can be used in imaging data to identify lenses. In galaxy-galaxy lenses, the background source forms a recognisable arc-like structure, often characterised by blue colors. In the case of point-like sources such as quasars, simply}

\subsubsection{The brightest objects} \label{magnification_bias}
Simply obtaining high-resolution imaging of the brightest quasars was an early strategic search for lensed quasars above $z\approx1$ \citep[for example with \textit{HST};][]{bahcall1992, maoz1992}, and is still a potentially fruitful method for the brightest high-redshift quasars \citep{mcgreer2013, fan2019}. The bright quasar catalogues of the Hamburg Quasar and Hamburg/ESO surveys have continued to provide small-separation lensed quasars even long after their initial identification \citep[e.g.][]{wisotzki1996, blackburne2008}. This increased fraction of lensed sources at the brightest end of the observed luminosity function is even stronger for sources with particularly steep luminosity functions -- namely very few bright sources and many faint sources. This is apparent at submillimeter wavelengths with dusty star-forming galaxies (DSFGs) \citep{coppin06}. \citet{negrello2010} took advantage of this by selecting bright sources within the 550 deg$^2$ of the Herschel Astrophysical Terahertz Large Area Survey (H-ATLAS), providing both a large enough area and high enough resolution to select five new lensed dusty star-forming galaxies without contamination from unlensed DSFGs, but still requiring low-redshift interlopers to be removed through colour cuts. Extensions of this technique have been used to build sizeable samples of strongly lensed DSFGs from the Herschel and South Pole Telescope surveys \citep[e.g.,][]{vieira2013,wardlow2013}, and to carry out the selection at even brighter magnitude limits across wider areas, using the Planck all-sky survey \citep[e.g.,][]{canameras2015}.

\subsubsection{Outliers from physical relations} \label{physical_relations}
The associated magnification of strong lensing can be used to identify outliers from non-linear physical relations. The brightest sub-mm galaxies (Section \ref{magnification_bias}) can be thought of as outliers from an underlying true luminosity function, however in this section we highlight objects with other physical properties or relations that are disobeyed when magnification changes the observed brightness.

\begin{figure*}[htb]
\centering
\includegraphics[width=\textwidth]{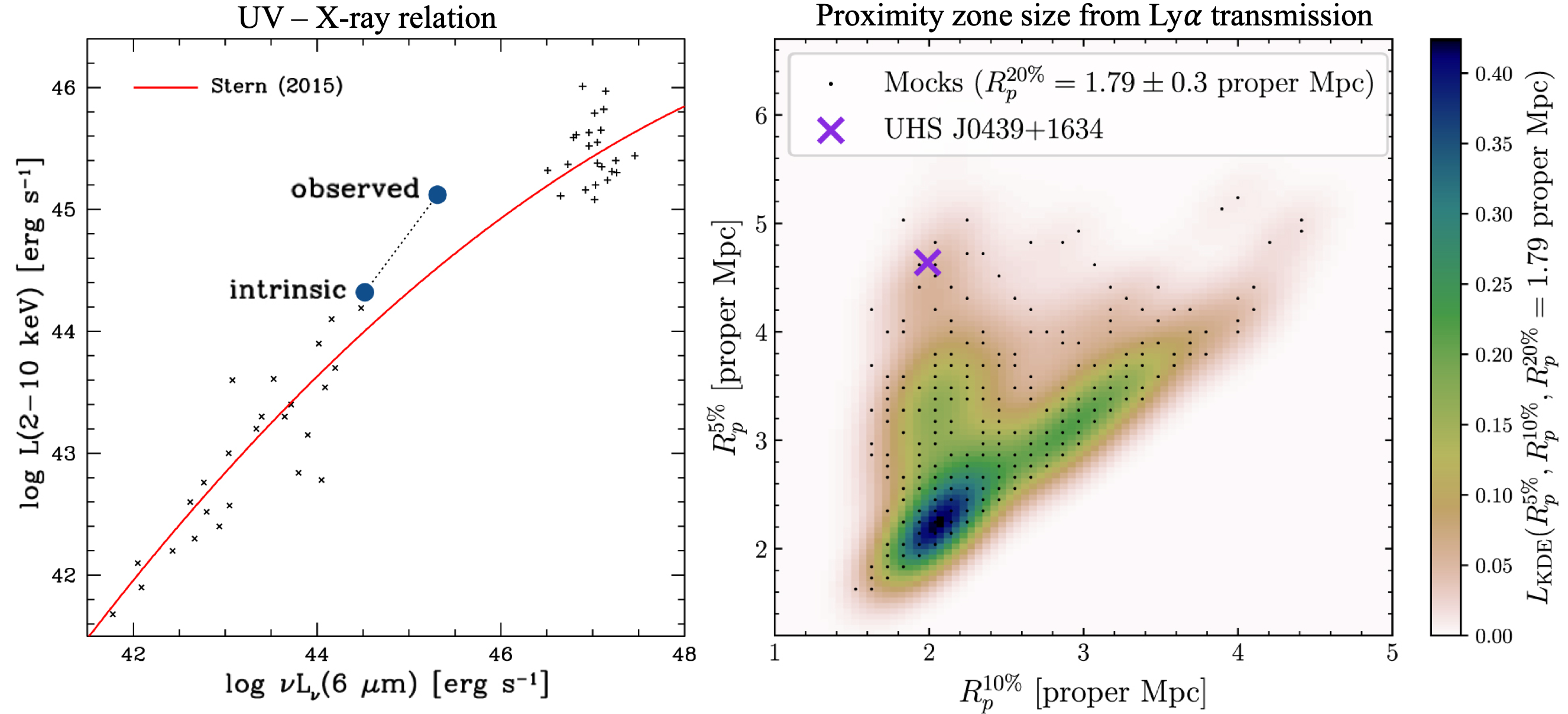}
\caption{Examples of lens discovery as outliers from physical relations, which can otherwise be explained by an associated magnification from lensing. \textit{Left}: MG1131+0456 UV - X-ray luminosity before and after the magnification against the observed relation, reproduced from \citet{stern2020}. \textit{Right}: proximity zone size for the lensed quasar UHSJ0439+1634 appearing as an outlier among mock systems with similar \textit{apparent} luminosities, reproduced from \citet{davies2020}.}
\label{fig:magnification_examples} 
\end{figure*}

The X-ray - UV luminosity relation for quasars has a clear flattening at higher luminosities. As outlined in \citet{stern2020}, if a source is strongly lensed, the UV and X-ray luminosities are increased by the same factor, taking them above the known relation, suggestive of lensing, as they demonstrated for the system MG1131+0456 (see Figure \ref{fig:magnification_examples}). It has also been used to argue against strong lensing for high-redshift lensed quasar candidates \citep{connor2021}: the physical size of the proximity zone of a quasar at high-redshift should be related to the intrinsic luminosity of the ionising quasar source. However, strong gravitational lensing will change the inferred luminosity, while the inferred size remains the same, and a discrepancy between the size-luminosity relation could support the lensing hypothesis. \citet{davies2020} showed that this technique can successfully recover the known $z=6.5$ lensed quasar, UHSJ0439+1634, while it conclusively rules out lensing for the candidate SDSSJ0100+2802 (see Figure \ref{fig:magnification_examples}). 

Another physical limit is that of accretion rates onto SMBHs, i.e. an Eddington ratio of unity. This has been used to propose an associated magnification to a quasar, in the case of J2329--0522 at $z=4.85$ \citep{yue2023}. In this case, however, no multiple imaging has been observed with \textit{HST} data, and thus has been dubbed an intermediately-lensed system, where the weak lensing magnification is still significant.

The magnification associated with lensing is most apparent when the source has a specific absolute luminosity. This is the case with type Ia supernovae -- they are \textit{standard candles}, all having similar peak absolute brightnesses. This feature is notably used to probe cosmological parameters, as the redshift and luminosity distances can both be measured. Assuming a cosmology, simply measuring the redshift of a type Ia supernova allows the prediction of the observed peak brightness. In the case of gravitational lensing, the additional magnification will lead to an observed peak brightness brighter than predicted. This is how the first galaxy-scale lensed supernova, iPTF16geu, was discovered \citep{goobar2017}. By obtaining spectroscopic redshifts and classifications of supernovae discovered in the intermediate Palomar Transient Factory (iPTF), iPTF16geu -- at z=0.409 -- was found to be 4 magnitudes (a factor of 50) too bright. High-resolution follow-up imaging clearly revealed the 4 multiple images, with a fraction of this extra magnification associated to microlensing \citep[e.g.,][]{more2017}. The discovery spectra also showed clear absorption signatures associated to the lensing galaxy, presenting a further sign of strong lensing. A second example, SN Zwicky, has also been found in similar fashion \citep{goobar2022}, with an Einstein radius below 0.2\arcsec, demonstrating that this selection technique can uniquely probe the full mass range of lensing galaxies. Even without a spectroscopic redshift, \citet{quimby2014} show how irregular SN Ia colours relative to observed magnitude can be used to infer a higher redshift and hence associated magnification.

Requiring a spectroscopic redshift creates a strong dependence on magnitude, as the brightest supernovae candidates are followed up preferentially. For current imaging surveys -- such as the Zwicky Transient Facility -- very few lensed supernova are expected to be brighter than this magnitude cutoff, due to the cross-section to lensing being small at low redshift. Future surveys, such as the LSST, will discover several hundreds of lensed supernovae. However, many of these cases will not require spectra for confirmation, since the LSST imaging quality will be able to resolve individual images. \citep{goldstein2017} proposed a method that relies neither on resolving the multiple images, nor a spectroscopic redshift, nor a classification of the supernova: by considering supernovae near ellipticals to be hosted by them, one can calculate the apparent magnitude if the supernova were type Ia using the photometric redshift of the elliptical. Since the vast majority of other supernovae are fainter than type Ia, a surplus in observed flux should then be associated to lensing magnification of a source supernova (not necessarily even type Ia when the magnification is large). 

\subsection{Multiple-imaging selection} \label{imaging_section}
\begin{figure*}[htb]
\centering
\includegraphics[width=\textwidth]{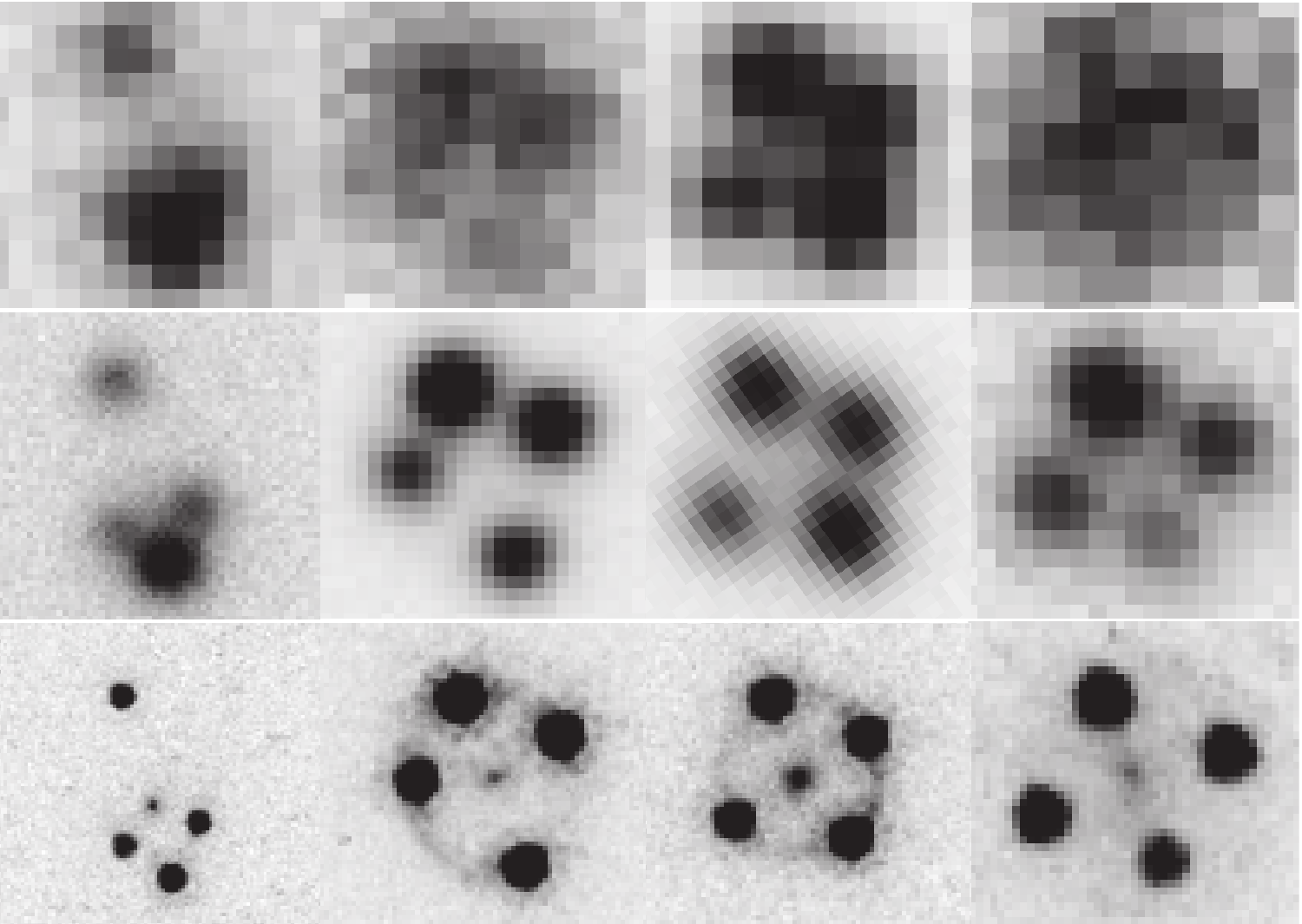}
\caption{\textit{Top}: Discovery images of four lensed quasars -- from left to right: DESJ0029--3814, ATLASJ0259--1608, DESJ0405--3308, and ATLASJ2344--3056. The pixel sizes of 0.26\arcsec\ (DES) and 0.21\arcsec\ (ATLAS) are typical of current ground-based surveys. \textit{Middle}: Magellan images obtained with pixel sizes of 0.07\arcsec\ (left three) and 0.11\arcsec\ (right) in excellent seeing.  \textit{Bottom}: \textit{HST} F814W imaging.}
\label{fig:lensedquasars}    
\end{figure*}

In this section, we discuss the techniques that identify lenses as systems containing spatially separated components as potential images of a background source. Typically this requires the analysis of optical and infrared imaging data (i.e., pixels), but in some cases catalogued detections can be used directly. In most cases, the starting data for a search are derived from wide-field ground-based imaging surveys, and higher resolution follow-up data are needed to not only confirm the lens nature, but also for most scientific analyses. Figure \ref{fig:lensedquasars} shows the difference in imaging quality between discovery data and subsequent higher-resolution imaging for four quadruply imaged lensed quasars.

\subsubsection{Citizen Science} \label{citizen}
Citizen science is one of the novel techniques that has emerged in the last two decades as a systematic means of doing science in Astronomy. Typically, it involves citizens from all over the world collecting and/or analysing data to collectively achieve a scientific goal. Repetitive tasks requiring special skills, such as pattern recognition, and that are difficult to carry out with computer algorithms are deemed suitable for citizen science. One of the first citizen science projects, Galaxy Zoo, was successfully used to study the morphology of galaxies \citep{lintott2008}.  Another major citizen science initiative -- The Zooniverse \citep{simpson2014} -- was setup to support many more citizen science projects, both within and outside of astronomy, including Planet Hunters, Bat Detectives, and Weather Rescue. Searching for gravitational lenses in modern wide-field, deep, multi-band imaging datasets is well suited to citizen science. %  {\bf ADD MORE and REFs}.

%As already discussed, the variety of structure, size, shape, colour and luminosity of strong lenses makes a complete and pure discovery algorithm extremely challenging. While visual inspection is indeed used by lens experts to ascertain the lensing nature, it is impossible to do this routinely at large scales owing to increasing numbers and sizes of astronomical surveys. 

Space Warps \citep{marshall2016} was the first citizen science project designed specifically to discover strong lenses from astronomical imaging surveys. The classification interface has tools to aid visual analysis, a spotter's guide which displays typical examples of lenses and non-lenses and the data stream is interspersed with training images unknown to the users. On the backend, a Bayesian pipeline combines classifications from multiple users to assign probabilities for an image to contain a lens. The end product is not only a high-confidence catalogue of lens candidates with calibrated probabilities, but also a vetted sample of contaminants which can be used by the community to improve their lens search algorithms, and for specific studies of spiral galaxies, red star-forming galaxies or ring galaxies.  

Space Warps has to-date conducted lens searches in VISTA-CFHT Stripe 82 Survey \citep{geach2015}, the CFHT Legacy Survey \citep{more2016} and the Hyper Suprime-Cam Survey \citep{sonnenfeld2020}. A unique feature of Space Warps is the supervised learning approach implemented in the analysis of the image classification wherein realistic training images are used. As a result, the final sample of candidate lenses can be well-characterised in terms of completeness and purity which is useful for the statistical analyses of lenses. Recently, a byproduct of the Zooniverse Hubble Asteroid Hunter project has been to identify strong lenses in archival \textit{HST} observations, identifying $\sim$200 new high-quality candidates \citep{garvin22}. Citizen science will remain an important way to test automated lens discovery algorithms in future surveys, such as LSST and \textit{Euclid}.

\subsubsection{Visual inspection}

Before citizen science projects, the brute force approach of inspecting all available imaging to look for lenses was only possible with small area surveys. In particular \textit{HST} has been extensively searched thanks to its depth and data quality, in the Hubble Deep Field \citep{hogg1996,barkana1999}, GOODS/ACS fields \citep{fassnacht2004}, the COSMOS fields \citep{faure2008,jackson2008}, the fields of known lenses \citep{fassnacht2006, newton2009}, and various other fields \citep[e.g.,][]{moustakas2007,more2011}. High-resolution imaging can also be obtained at radio wavelengths: the Cosmic Lens All-Sky Survey (CLASS) imaged over 13,500 radio sources at a resolution of 0.2\arcsec\ with the Very Large Array (VLA) at 8.4 GHz \citep{myers2003, browne2003}. Together with the $\sim$2400 images from the Jodrell Bank VLA Astrometic Survey \citep[JVAS,][]{king1999}, the sample was inspected for systems with multiple components, and promising candidates were followed-up for higher-resolution imaging with the Multi-Element Radio Linked Interferometer Network (MERLIN) and Very Long Baseline Array (VLBA), each at 5GHz. These efforts led to the discovery of 22 new lensed quasars. Similar investigations have been carried out with the VLA in the Southern hemisphere, which uncovered 4 further radio-loud lensed quasars \citep[e.g.,][]{winn2002}, and again more recently with the VLBA using a different parent sample \citep{spingola2019}. 

In lower-resolution imaging, the lensing nature is often not obvious, except for the widest separation systems. The 14,000 square degrees of SDSS imaging fuelled a variety of such searches for wider-separation lensed galaxies (typically with images separated by $>$3\arcsec\ from the lensing galaxy), through visual inspection of red galaxies with nearby blue catalogued detections \citep[e.g.,][]{shin2008, kubo2009, diehl2009, stark2013}. Before visual inspection, images can be pre-processed to remove the bright lensing galaxies that often hide faint background arcs, either by simply subtracting a rotated version of the system \citep{anguita2012}, or through Principal Component Analysis \citep[][]{joseph2014,paraficz2016}. Similarly, giant arc searches have been performed by visual inspection of pre-identified clusters in both \textit{HST} \citep{sand2005}, SDSS \citep[e.g.,][]{estrada2007, hennawi2008}, the Dark Energy Survey \citep[DES,][]{diehl2017, odonnell2022}, and at group-to-cluster scales in the Hyper Suprime Cam Subaru Strategic Program Survey \citep[HSC-SSP,][]{jaelani2020}.

However, for those lenses with smaller separations, comparable to the Point Spread Function (PSF) width of the survey, contaminant systems resembling lenses quickly become overwhelming and outnumber the real lenses. For lensed quasars, these contaminants are most often quasars projected close to foreground stars, or compact star-forming galaxies. Thus, such searches are designed at specifically removing these contaminants. % by combining imaging with other methods based on photometry, catalogues, or machine learning techniques.

In the remainder of this section, we will discuss techniques that apply some algorithms for reducing the number of candidates to a small fraction of the input catalogue. However, these searches still require a final step of visual inspection of cutouts, sometimes many tens of thousands. Biases from individual inspector's subjectivity can naturally lead to biases towards certain types of lenses (see Section \ref{sec5:bias}), and it is common practice to have multiple authors inspect the same stamps, or inject true positives into the sample to assess this bias statistically. \citet{rojas2023} investigated the effects of visual classification by using a labelled training set of 1500 mock and real images of lenses and non-lenses. 55 classifiers with varying levels of experience in lens discovery graded the systems through visual inspection of their colour images. As expected, arcs with low signal-to-noise or Einstein radii less than or equal to the PSF width are rarely recovered. They found that substantial variations are found between individual classifiers which can be mitigated by combining scores from 6 or more individual classifiers.

\subsubsection{Catalogue searches} 
\label{catalogue_selection}
Wide-field imaging and spectroscopic surveys are now available across various wavelengths, meaning that all-sky classification and cataloguing of galaxies and quasars is possible down to a few arcseconds resolution. Simple queries and self-cross-matches within these catalogues can be used to look for nearby sources with similar colours as lens candidates. One of the first dedicated catalogue-based lens searches of this type was the SDSS Quasar Lens Search (SQLS), which obtained deeper imaging and/or long-slit spectroscopy of spectroscopically confirmed quasars from SDSS which were either extended or had multiple nearby components in the SDSS imaging catalogue. Colour cuts on companions reduced the many contaminants, and the multi-year campaign discovered 49 new lensed quasars \citep{oguri2006, inada2012}. These searches have been extended by including additional infra-red imaging for candidate selection \citep{jackson2012}, or simply by application to newer spectroscopic samples \citep{more2016b}.

Lensed quasars can also be discovered as outliers from the unlensed quasar population in their catalogued magnitudes, due to the additional contribution from the lens galaxy, which leads to an apparent near-infrared flux excess. \citet{ofek2007} applied this colour cut on SDSS spectroscopic quasars with red $g-H$ colours using the Two Micron All Sky Survey (2MASS), successfully identifying a new double quasar.

Even without spectra, quasars can be selected photometrically thanks to their unique infrared colours with the Wide-field Infrared Survey Explorer all-sky survey \citep[WISE,][]{wright2010, stern2012}, and various radio and X-ray surveys \citep[see, e.g.,][]{flesch2021}. Searches using WISE photometric quasar candidates coupled with resolved or extended detections in ground-based imaging data such as DES and VST-ATLAS have yielded a handful of new lenses \citep[e.g., ][]{agnello2015, ostrovski2017}, but typically with significant contamination from quasar-star projections and star-forming galaxies which mimic the infrared colours of quasars \citep[e.g.,][]{williams2018}. WISE alone is limited due to its resolution of $\sim$5\arcsec, resulting in the photometry of any quasar-star projection becoming blended and appearing very similar to that of a lensed quasar. Significant visual inspection is often required as a final step in these processes \citep{rusu2019, dawes22, he2023}.

The high level of contamination in photometric searches was partially addressed by \textit{Gaia}, which provided an all-sky, space-based-resolution map of bright point sources \citep{gaia}. For known lens systems, the multiple lensed images were often all catalogued in \textit{Gaia}, and contaminant systems could be removed with various catalogue parameters \citep[e.g.,][]{lemon2018}. Cross-matches between the aforementioned photometric quasar catalogues and \textit{Gaia} yielded several dozen new lenses \citep{agnello2018, Spiniello18, desira2022}, aided by the proper motion and parallax parameters and improved completeness in the second \textit{Gaia} data release. For those systems with only one image bright enough to be detected or simply with undetected images, the single \textit{Gaia} detection was useful when combined with ground-based imaging. By looking for astrometric offsets between the resolved \textit{Gaia} detection of one image and the blended weighted centroid of the ground-based image, coupled with a difference in the overall photometry, \citet{lemon2017} discovered several candidate lensed quasars which were otherwise only seen as single point sources in ground-based imaging. This is similar to the method employed by \citet{jackson2007}, who looked for offsets between radio positions from FIRST and optical SDSS galaxy positions. Methods making use of \textit{Gaia} data alone and its improved completeness and various catalogued parameters has unveiled a potentially large population of new lenses and dual quasar candidates, thanks to high-resolution \textit{HST} follow-up imaging  \citep{shen2021, chen2022, mannucci2022}. However, contamination from quasar-star projections still dominates, as detections with close neighbours either do not have recorded proper motions or parallax, or these are consistent with zero for most faint stars. Simultaneous use of \textit{Gaia} and pixel-fitting of other imaging surveys has overcome this remaining contamination (Section \ref{pixel}).
%Although expensive, adaptive-optics imaging on 10m-class telescopes is a powerful way to discover lensed quasars within these quasar pair samples by revealing the lensing galaxy that is often hidden by the bright quasars in seeing-limited observations \citep{shajib2021b}.
Quadruply imaged point-like sources occupy specific regions in the position/flux-ratio space, such that any experienced observer can easily recognize lens systems from spurious star/galaxy asterisms by only looking at the image positions and their relative magnifications. Indeed, depending on the position of the source compared to the tangential caustics, one usually ends up with three type of configurations: (i) cusps, when the sources stands near the intersection of two caustics, (ii) folds, when close to a single caustic and (iii) crosses, when the source lies close to the centre of the caustics. 

The position of four postulated images can be forward modelled by a simple lensing potential \citep[e.g.,][]{falor2022}, such as the Singular Isothermal Ellipsoid Potential (SIEP), which often recovers image positions for most known lensed quasars, and requires seven free parameters for the fit (2 for the source position, 2 for the galaxy position, and 3 for galaxy parameters, i.e., Einstein radius and shape). \citet{witt96} demonstrated that the galaxy, source, and images of a SIEP must all lie on a hyperbola with asymptotes along the major and minor axes of the potential. This reduces the dimensionality of the model to three degrees of freedom, while the constraints are simply the positions of the four images. \citet{wynne18} also shows that the images of an SIEP must lie on an ellipse centred on the source and axes parallel to the hyperbola's asymptotes. They subsequently suggest a two-dimensional minimisation  (source position along the hyperbola and potential axis ratio) to find the best fit model, from which one can report a figure of merit which determines the disagreement from a SIEP model. They report that for a certain threshold of this figure of merit, 98\% of random quartets can be removed while only removing 20\% of known quad quasars -- namely those with peculiar potentials such as double lensing galaxies. \citet{schechter19} present an even faster determination of the best-fit model by insisting that the two brightest images fall on a shifted copy of the Witt hyperbola, ensuring always a four-image solution and no need for solving a minimization problem. 

The application to lens finding simply requires four positions, which promises an extremely quick lens discovery technique in the catalogues of high-resolution imaging surveys. Indeed, this method recovers the majority of known lensed quasars from \textit{Gaia}, suggesting that there are no bright missing quads. Machine learning has also been applied to this problem, and is discussed in Section \ref{machine_learning_catalogues}.

We conclude our review of catalogue-based methods with cluster lensed quasars. Typically, giant arcs in clusters can be difficult to detect due to their low signal-to-noise and require pixel-level search algorithms (see Section \ref{pixel}). Quasar sources, however, are often bright enough to be available in ground-based imaging catalogues. \citet{Shu2019} cross-matched their \textit{Gaia} AGN catalogue to known bright cluster galaxies, and discovered a new 21\arcsec\-separation doubly lensed quasar. Similarly \citet{yantovskibarth2023} created a catalogue of clusters from the Legacy Surveys imaging, and found 5 new high-grade lens candidates (of which 2 were known) by searching for multiple DESI quasar targets within the Einstein radius of their clusters.

\subsubsection{Pixel-level searches} \label{pixel}

%\chiara{Moreover, matters are further complicated by the  atmospheric differential refraction (ADR) and the choice of different astrometric calibrators for each survey. \citet{agnello2018} resolved this by means of field-corrected offsets, using the 'Blue and Red Offsets of Quasars and Extragalactic Sources' code\footnote{BaROQuES, available upon request, \url{https://github.com/aagnello}.}}.

For lensed quasar discovery, the astrometric lensing information is entirely in two (doubles) or four (quads) relative image positions (often without a lensing galaxy visible). Doubles are often indistinguishable from common contaminant quasar-star or star-star pairs, and thus careful pixel-level analysis that looks for colour similarity through the use of multiple survey datasets remains the leading discovery methodology. For lensed galaxy discovery however, the state-of-the-art is machine learning. We defer the discussion of these searches to their own section (see Section \ref{machine_learning}), but here describe the earlier techniques that were applied to look for arcs. 

A major limitation of catalogue searches in ground-based imaging datasets is related to deblending -- for any lens system with images separated by less than the size of the PSF, the separate images and lensing galaxy risk being catalogued as a single object. Other methods, such as identifying multiple redshifts in spectra \citep[e.g.,][see Section \ref{fibrespectra}]{Bolton2004}, offer ways to overcome this limitation, however it is still possible to detect the multiple images by analysing the pixels themselves. \citet{schechter2017} fit multiple components to the multi-band pixels of VST-ATLAS for photometric quasar candidates, prioritising systems containing components with similar quasar-like colours, discovering four quasar pairs and three lensed quasars. They remove star-forming galaxies, a common source of contamination for WISE-selected photometric quasars, as systems with components incompatible with a given image's PSF. This technique has also been extended to DES pixels \citep{anguita2018}. \citet{chan22} have applied the same principle to the 0.6\arcsec\  median-seeing survey of the Ultraviolet Near Infrared Optical Northern Survey (UNIONS), identifying nearby point sources with similar $u-r$ colours via a convolution of the Laplacian of the point spread function.

Given the precise positions of the \textit{Gaia} detections, \citet{lemon2019} analysed the pixels of WISE, to extract the photometry at the known positions of the individual point sources from \textit{Gaia}. Simple colour cuts on the two components were able to remove 80\% of previously identified quasar/star projections and retain 99\% of lensed quasars. Applying this selection, alongside inspection of Pan-STARRS and DECaLS imaging, long-slit spectroscopic follow-up for 208 candidates was obtained with 2m- to 4m-class telescopes (ISIS on the WHT, ALFOSC on the NOT, and EFOSC2 on the NTT), with a success rate (quasar pairs or lenses) of 88 \% \citep{lemon2019, lemon2023}. \textit{Gaia}'s limiting magnitude is $\sim$21, while ground-based imaging surveys are now reaching beyond this, and should resolve many fainter lensed quasars.

Pixel-level analysis for compatibility with lens configurations has also been used for final candidate selection. Typical features in lensed galaxies at both galaxy- and cluster-scales are arcs, which can be targeted through algorithms built to identify such elongated and curved features. Several `arc-enchancing' algorithms have been published, which simply require detection of significant structures in the resulting images: \citet{horesh2005} targeted giant arcs in clusters by applying \textsc{Sextractor} multiple times, repeatedly removing non-elliptical sources from the segmentation map; \citet{lenzen2004} applied anisotropic diffusion to enhance arcs along their direction while keeping them the same thickness and reducing noise; and \citet{alard2006} defined a local elongation map based on second order moments within a few times the effective PSF size. The latter was successfully applied to the Canada-France-Hawaii Telescope Legacy Survey (CFHTLS) by \citet{more2012}. \citet{seidel2007} similarly use moments to determine local elongation but in cells of a grid, which are subsequently grouped with nearby cells sharing similar local elongation directions, resulting in groups of arc-like structures which can have curvature, and do not require a subsequent detection algorithm. This arcfinder has since been extended to require colour information for the arcs \citep{maturi2014}. \citet{gavazzi14} presented and applied their \textsc{RingFinder} algorithm to CFHTLS; by analysing PSF-matched $g-\alpha i$ images of galaxies (i.e., scaled $i$-band images subtracted from $g$-band images to remove the galaxy light), \textsc{RingFinder} finds connected tangential pixels between 0.5\arcsec\ and 2.7\arcsec\ from the galaxy, requiring at least two detections or one elongated detection. \citet{lee2017} use a circular Hough transform to identify overdensities of light lying on rings in SDSS LRG cutouts, recovering the four known Einstein rings in the sample, with 40\% purity. 

Requiring a more strict compatibility with a source lensed by a simple mass model has been applied at the full pixel-level light distribution \citep{marshall2009} and on extracted image positions \citep[\textsc{Chitah}, ][]{chan15, sonnenfeld2018, chan2020}, which results in high purities, but at the risk of missing systems with more complex lensing potentials. \textsc{YattaLens} combines arc detection and lens modelling for lens discovery \citep{sonnenfeld2018}. Arc candidates are first found as tangential detections in galaxy-subtracted $g$-band images of the HSC-SSP survey; subsequently the pixels are fit with a lens model (Sersic for lensing galaxy + lensed Sersic for arcs) and with a non-lens model consisting of only Sersic profiles. The systems with a better fit from a lens model were kept as candidates. The algorithm has been further applied to find new lenses in more recent data releases of HSP-SSP \citep{sonnenfeld2020, wong2022}.

\subsubsection{Variability} \label{variability}
\begin{figure*}[htb]
\centering
\includegraphics[trim={1cm 2.8cm 1cm 3.8cm},clip,width=\textwidth]{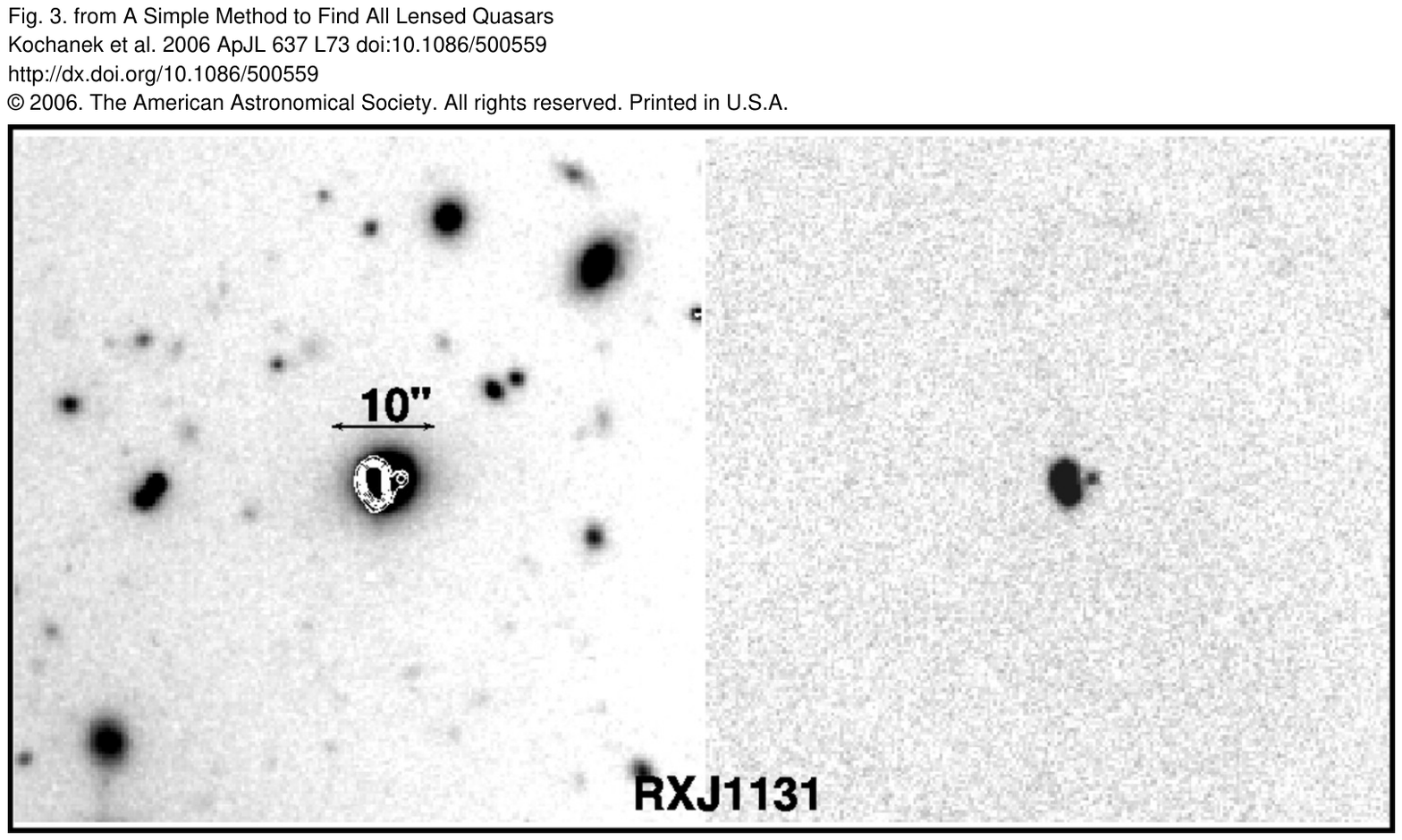}
\caption{Figure 3 from \citet{kochanek2006} showing the field of RXJ1131-1231 (left) and the variability map built from several epochs (right). The lens system is clearly the only extended variable source in the field.}
\label{fig:variabilitymap}    
\end{figure*}

In recent years, cadenced wide-field optical imaging surveys have opened up the possibility of finding variables and transients by comparing different epochs of imaging, via a method called \textit{difference imaging}. If the source quasars are sufficiently variable, lensed quasars appear as unique sources in such difference images, since they consist of multiple crowded variable point sources, unlike most variable objects, as first proposed by \citet{kochanek2006}. They predicted that the dominant source of extended variable objects as seen by examining combined difference images would be lensed quasars, with some small contamination from binary quasars, quasar/variable star projections and variable star pairs. Figure \ref{fig:variabilitymap} shows such an extended stacked variability map of a known lensed quasar. \citet{lacki2009} first demonstrated the efficacy of this technique on 20,536 candidates in the SDSS Supernova Survey region, finding only eight extended variables recovering the only known lensed quasar in the sample. \citet{zuzanna2018} searched 670 square degrees of OGLE (Optical Gravitational Lensing Experiment; an optical imaging survey of the Magellanic Clouds region with cadence of $\sim$3 days) for red WISE sources, i.e. photometric quasar candidates, with at least two variable components within 6 arcseconds yielding 63 candidates, of which 3 showed similar lightcurves. Two were spectroscopically confirmed as quasar-star projections, while OGLEJ0218-7335 was confirmed as a new lensed quasar. Within this area and to the single epoch depth of OGLE, approximately 10 lensed quasars are expected. \citet{lemon2020} used component modelling of DES epoch data to measure variability properties of lensed quasars and quasar-star projections, further demonstrating variability as a powerful selection and prioritisation tool. They also uncovered a \textit{variability bias} associated with finding lensed quasars through variability alone, further discussed in Section \ref{variabilitybias}. More recently attempts have been made to implement such a variability search in HSC by \citet{chao2021}, however the implementation of the difference imaging and selection criteria can limit the purity and efficiency of such a method, and many artefacts can lead to large numbers of false positives. \citet{dux2023} use the multiple epochs of Pan-STARRS and apply difference imaging on close \textit{Gaia} pairs with quasar-like infrared colours in WISE; they present spectroscopic follow-up with only 25\% contamination from stellar systems.

We note that such a search could be applied to any lensed transient in which the image separation is greater than the resolving ability of the imaging survey, for example with the discovery of lensed supernovae in the optical, however, this technique would typically be complemented with other methods to ensure earlier detection. Finally, we note that as long as the time delay between images is much shorter than the survey length, each image should have approximately similar variability (if ignoring microlensing). Indeed a non-zero time delay is not necessary for this technique to work, in contrast to the discovery methods outlined in Section \ref{temporal_selection}.

%%%%
\subsection{Multiple-redshift selection}
In this section, we discuss the techniques that identify lenses from spectra exhibiting features at multiple redshifts. The Einstein cross was the first spectroscopic lens discovery \citep{huchra1985}, thanks to the serendipitous detection of broad emission lines in a galaxy spectrum during a redshift survey of nearby galaxies. Despite no clear signs of the multiple images, the presence of a bright $z=1.7$ quasar in the spectrum of a $z=0.0396$ bright spiral galaxy led to the likely conclusion of strong gravitational lensing. In most cases, fibre spectra have led to the identification of lens candidates in exactly the same way: identifying a higher redshift galaxy in the spectra of lower redshift galaxies. This technique can be extended to resolved spectroscopy at a smaller scale, i.e. Integral Field Units (IFUs), while overcoming the drawbacks of fibre spectra.

\subsubsection{Fibre spectra} \label{fibrespectra}
\begin{figure*}[htb]
\centering
\includegraphics[width=\textwidth]{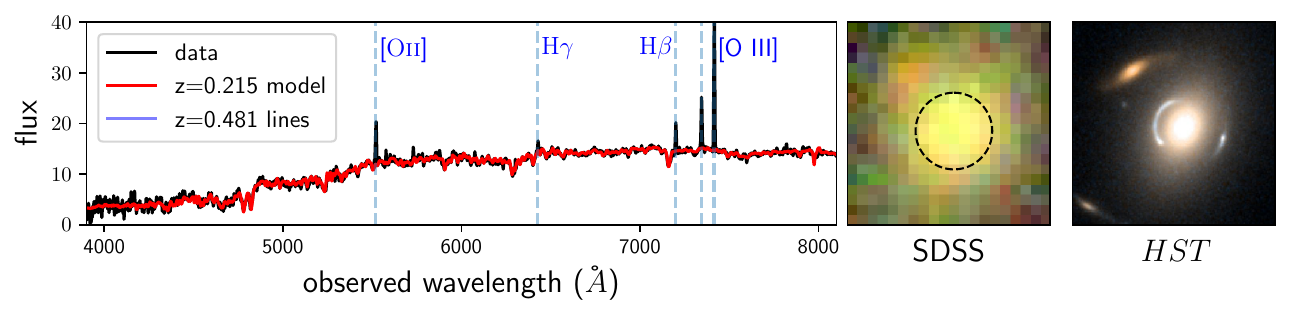}
\caption{SDSSJ1205+4910 discovered by \citet{bolton2008}. \textit{Left}: an SDSS spectrum of a $z=0.215$ LRG shows additional emission lines not fit by the model template, but that of a higher-redshift source at $z=0.481$; \textit{middle}: SDSS \textit{grz} imaging with fibre overlaid; \textit{right}: \textit{HST} ACS F555W, F814W colour composite shows clear signs of lensing. }
\label{fig:emissionlines}
\end{figure*}

In the early 2000s, the arrival of the first SDSS data releases provided hundreds of thousands of high-quality spectra for targeted searches of background source features. \citet{Bolton2004} searched 50,000 LRG spectra for higher redshift oxygen and hydrogen nebular emission lines, finding 50 high-quality candidates which became the basis of the Sloan Lens ACS Survey (SLACS) -- an \textit{HST} snapshot imaging survey for strongly lensed galaxies. Figure \ref{fig:emissionlines} shows an example of such a lens discovered by SLACS. The efficiency of requiring multiple emission lines, well-fit LRG spectra, and larger Einstein radii led to a final sample of 70 definite strong lens systems \citep{bolton2008}. This selection has since been repeated many times, with new datasets \citep{holwerda2015,talbot2021} or machine learning selection \citep{li2019}, and targeting specific lenses and/or source types: edge-on late-type lenses \citep{Treu2011}, higher-redshift \citep{Brownstein2012} and lower-mass lenses \citep{Shu2015}, and higher-redshift sources \citep{Shu2016, cao2020} or early-type sources \citep{Auger2011, Oldham2017}. Quasars acting as lenses have been found in a similar way \citep{courbin2012, meyer2019}, and examples of lensed quasars with galaxy-quasar blended SDSS spectra have been found \citep{lemon2019}.
 
Due to increased fibre numbers, telescope sizes, and instrumentation improvements, next-generation spectroscopic surveys, such as the Dark Energy Spectroscopic Instrument (DESI), will obtain spectra for tens of millions of galaxies, allowing for the discovery of lensed galaxies and quasars with fainter, higher-redshift lensing galaxies.

\subsubsection{IFU spectra}
While fibre spectra can efficiently provide spectra of many galaxies, they provide only integrated light within the typically 3\arcsec-wide fibres. This leads to two major drawbacks for lens searches: (i) if the lensed images lie outside of the fibre, the object will not be selected, and (ii) high-resolution follow-up imaging is almost always required to rule out a wide separation projection without evidence of strong lensing. On the other hand, imaging-only searches (Section \ref{imaging_section}) can immediately confirm the lensing nature, but follow-up spectroscopy is always required to convert lensing values into physical quantities. IFUs overcome these drawbacks, by simultaneously providing both spatial and spectral information. It is unsurprising that IFUs with wide fields of view have thus been used for targeted lens searches. 

Targeted IFU observations have been used to locate the rare low-redshift lenses -- particularly powerful probes of measuring the stellar mass-to-light ratio and hence the Initial Mass Function (IMF) due to probing the lens at smaller radii (where dark matter can be neglected, and thus lensing probes only the stars). The SINFONI Nearby Elliptical Lens Locator Survey \citep[SNELLS,][]{Smith2015} surveyed 27 galaxies selected to have the largest velocity dispersions (and thus largest lensing probability), with observations in the near-infrared to optimise the chances of observing lensed H$\alpha$, discovering two new lenses. This technique has been repeated with SDSS-MANGA \citep{smith2017,Talbot2018}, and MUSE with new targets and archival observations \citep{collier2018,Collier2020}. Even those systems with only singly imaged nearby background objects provide competitive constraints on the IMF of nearby galaxies. Future surveys, such as Hector, will provide targeted narrow-field IFU observations of over 50,000 galaxies, and will thus offer a chance of creating larger low-redshift lens samples. The ever-increasing area of archival wide-field IFU observations also promises further lens discoveries of various source and lens types \citep[e.g.,][]{galbany18}.

\subsection{Temporal selection} \label{temporal_selection}
A unique feature of gravitational lenses is their associated time delay. When the source is variable, this time delay can be measured. However, it can also be used to search for lenses with variable sources. While there is overlap with the idea of using variability of nearby objects (as outlined in Section \ref{variability}), we focus on lens searches that require a non-zero time delay. For the following techniques, the length of the dataset must be at least the length of the time delay, and the exact success rates will depend strongly on the sampling, signal-to-noise, and intrinsic source lightcurve shape. Typically, variations with large absolute second time derivatives are most useful for detecting a time delay.

\subsubsection{Lightcurves}
\citet{pindor2005} showed that resolved lightcurves can significantly improve a search for lensed quasars by targeting systems with well-detected time delays through the use of a dispersion statistic between pairs of lightcurves. Even earlier, \citet{geiger1996} proposed using autocorrelations of blended lightcurves to measure time delays in radio monitoring of lensed quasars, bypassing the need for expensive interferometric monitoring to resolve the multiple images. While their aim was not discovery, the principle has been recently revisited as a way to find lensed quasars in blended optical lightcurves, fueled by the promise of many hundreds of thousands of well-sampled quasar lightcurves in upcoming large imaging surveys. After subtracting a best-fit damped random walk (DRW) model, \citet{shu2021} found robust features in the autocorrelation function at the expected true time delay, with true positive rates of up to 50\% while keeping the false positive rate below 10\%. Application on real COSMOGRAIL lightcurves returned $\sim$ 20-25\% of the systems. \citet{bag2022} propose a method of reconstructing lightcurves and identifying lenses as those with minimum fluctuations at non-zero time delays. \citet{denissenya2022} have applied a similar technique to blended lensed supernovae using freeform templates. By considering the best-fit models for one, two, or four images, and using the Akaike Information Criterion, they find they can always recover the number of images for systems with modest magnification ratios and time delays longer than 10 days. \citet{kronemartins2019} note that the blended lightcurve is less stochastic than that of an unlensed quasar, and thus has lower entropy. Applying this low-entropy selection to quasar lightcurves in the Catalina Real-Time Transient Survey with multiple \textit{Gaia} detections led to the successful discovery of new doubly imaged lensed quasars.

\subsubsection{Transients} \label{transient_temporal}
Observations of GWs, GRBs, and FRBs often provide uniquely well-sampled time baselines that would allow for an exact time delay measurement if a source were strongly lensed into multiple images. The principle for detection of a lensed event is straightforward: for each image, all the parameters should be identical (except for the arrival time, luminosity distance, and coalescence phase). Therefore, a search for signals with similar parameters may highlight pairs of lensed images from the same source. For GWs, \citet{ligosearch2021} explore a ranking statistic for such a posterior overlap, and a joint posterior estimation between pairs of events. In most of the (circular, non-spinning) binary systems, we expect lensing to only change the amplitude of the signal and introduce a constant phase shift between a pair of lensed events. Thus, some methods simply search for similarity in the shape of the GW signals in the time and/or frequency dimensions as a means of identifying lensed counterparts. %, for instance, machine learning based lens searches.
Accounting for overlap of sky localisation regions, population priors, selection effects, and the odds against lensing, no convincing lens pairs have yet been found. We point the reader to the Gravitational Wave Chapter for a full description of lensed GW detection methods.

\smallskip
For GRBs, 6 pairs have been identified as possible image pairs of a single event. A summary of relevant properties is given in Table \ref{table:lensedgrb_summary}. A common feature of these candidate events is their short time delays. This has led the studies' authors to posit that their lenses are black holes in the range $10^4$ \(M_\odot\) < M < $10^6$ \(M_\odot\). %, with an optical depth to strong lensing comparable to that of the galaxies that lens quasars. 

\begin{table}[h!]
\caption{We list the main properties of paired GRBs: times between the pairs of events in seconds, the observed flux ratios (namely the measured luminosity of the first pulse divided by the second pulse), total magnification (summing both images) for a point mass lensing model, and the inferred lensing mass (including a factor of $1+z$, where $z$ is the unknown lens redshift).}
\centering
\begin{tabular}{c c c c c c}
 \hline
 Name & $\Delta$t (s) & ${\mu}_{1}$/${\mu}_{2}$ & $\mu_{\textrm{macro}}$ & $M_\textrm{lens}$/$M_\odot$ & Reference \\
 \hline
  GRB 950830 & 0.4 & 1.8 & 3.5 & $10^{4.4}$ & \citet{paynter21} \\
  GRB 200716C & 1.9 & 1.5 & 5 & $10^{5.3}$ & \citet{wang2021}\\
  GRB 200716C & $^{\prime\prime}$ & $^{\prime\prime}$ & $^{\prime\prime}$ & $^{\prime\prime}$ & \citet{yang21}\\
 GRB 210812A & 33 & 4.3 & 1.6 & $10^{6}$ & \citet{veres21} \\
 GRB 081126A & 31 & 1.35 & 6.7 & $10^{6.7}$ & \citet{lin22} \\
 GRB 090717A& 42 & 1.69 & 3.9 & $10^{6.6}$ & \citet{lin22} \\
 
\hline
\end{tabular}

\label{table:lensedgrb_summary}
\end{table}

\smallskip
FRB searches are again similar, but are constrained to detect time delays below the duration of the data dumped by detection telescopes (see Section \ref{sec5:bias}), corresponding to lens masses of $\sim10^{4} M_{\odot}$. Incoherent FRB lensing searches use detailed studies of individual lightcurve morphologies to distinguish between a doubly-imaged FRB and the more-common scenario of repeat emission from the same source \citep{liao2020constraints, krochek2022constraining, zhou2022search} -- approximately $5-10\%$ of FRB sources emit repeat bursts. These searches probe minimum time delays comparable to the pulse width. Another class of searches overcomes the degeneracy between the pulse morphology and gravitational lensing by relying on the compact emission region of the FRB and detecting the interference pattern between the two images in the time domain \citep{kader2022high, leung2022constraining}. These coherent searches are limited by the fundamental time resolution of the telescope, though radio scattering by inhomogeneous gas in foreground haloes remains an obstacle to detection. These complementary search methods probe different timescales and therefore different lens mass ranges, and upper limits on the fraction of dark matter comprised of compact lenses have been published, though detections of lensing have not yet been claimed.

%Supposing that these black holes were of baryonic origin, one might expect them to be concentrated inside the Einstein rings of their host galaxies, in which case one might expect additional images with weeks-long delays. Alternatively, if they are primordial, they would be distributed like the dark matter in the halos of galaxies, and would appear to violate the upper limits in the recent review by \citet{green2021}. Moreover, they might be subject to strong tides and hence quadruply rather than doubly imaged. 

\subsection{Machine Learning} \label{machine_learning}

%%% Added by LD on 2022-10-03 %%%
Machine learning is the field of artificial intelligence devoted to the development and study of methods that are able to `learn'. From a user-provided sample of training examples, these algorithms either:

\begin{itemize}
    \item build arbitrarily complex relations between input attributes and numerical or discrete output values that maximise a given score measure in \emph{supervised} learning methods; these models are then used on observations to perform predictions
    \item group observations according to the similarity of their input attributes according to a given similarity measurement in \emph{unsupervised} learning methods.
\end{itemize}

Machine learning dates back to the 1950s. Examples include teaching machines to play the game of Checkers \citep{Samuel59} and Rosenblatt's \textit{perceptron} -- a linear neural network capable of recognising letters. In the search for strong gravitational lenses, supervised machine learning methods are commonly used as two-class classification problems (lens vs non-lens) with output values representing the probability of the input being a lens. These classification techniques can be viewed as automatically drawing frontiers in a multidimensional input space of attributes in order to isolate objects of two intrinsically different types.  % The further these frontiers are from the observations, the lower the uncertainty are on the predictions.
%\subsubsection{Historical background [Ludovic]}
%%% Added by LD on 2022-01-12 %%%
%%% Completed by LD on 2022-10-03 %%%
Machine learning's first successful applications for lens finding were as late as 2017 \citep{Petrillo2017, Jacobs2017, Lanusse2018}, partly due to the advent of wide-field, deep surveys, but also to the requirement of large computational resources associated with the processing of astronomical images. A comparison of lens-finding algorithms, including conventional methods (e.g., arc finders and visual inspection), and machine learning methods, such as support vector machines \citep[SVM, e.g., ][]{hartley2017} and convolutional neural networks (CNNs), was presented in the `strong gravitational lens-finding challenge' \citep{Metcalf2019}.  %These resources were however not broadly available before 2010 and the emergence of GPU processing. C.L.: this sentence is already implied
%In 2016, a collaborative effort was undertaken in order to foster the development of new methods that can scale the rapidly increasing amount of data produced by upcoming surveys like the LSST or Euclid. This study, finally presented in \cite{Metcalf2019}, aimed to optimize the purity and completeness of galaxy-galaxy lens samples selected from mock images using various approaches such as visual inspection, arc detection algorithms, classical machine learning methods (Support Vector Machines and Logistic regression) and deep learning methods based on Convolutional Neural Networks. 
Here, we will focus on the main machine learning techniques that have been applied to lens searches: CNNs and decision trees, however the field is quickly evolving, with new techniques constantly being developed and applied to new datasets. We direct the reader to \cite{Bishop2006} for a more in-depth introduction to machine learning, and to \citet{huertas2023} for an application of machine learning to astronomical survey data (which includes an overview of its application to lens finding and lens modelling).

\subsubsection{CNNs}

Since the early efforts by \citet{Dieleman2015} and \citet{Huertas2015}, deep learning algorithms have become state-of-the-art methods for galaxy morphological classifications based on pixels. These techniques similarly offer a solution to extract the pixel-level information contained in strong gravitational lenses, especially for systems with extended, spatially-resolved lensed features. The machine-learning algorithms best optimised for image analysis are supervised CNNs \citep[][]{Lecun1998}. CNNs are able to capture the morphological patterns in imaging data by learning the coefficients of convolutional kernels and creating a range of two-dimensional feature maps. In practice, between the input layer of astronomical images, and the output layer of corresponding labels, CNN architectures comprise a range of convolutional, pooling, and fully connected hidden layers. To differentiate the network from a simple linear regression, non-linear activation functions, such as the hyperbolic tangent function, are inserted between each one of these layers.

Like all supervised learning methods, CNNs rely on a ground-truth dataset comprising images and their associated labels. During training, CNNs adjust their convolutional kernel weights by minimising a loss function, which encodes the difference between the ground truth and predicted labels. After each pass through the entire training set (refereed to as an epoch), the model is evaluated on a validation set to evaluate its performance on independent data. The evolution of training and validation losses are monitored, and the best CNN model with good generalisation to the validation set corresponds to the epoch with lowest validation loss. Information learned by the CNN is eventually stored in the feature maps. The size of labeled datasets required for training and validating CNNs for image classification vary between 10$^4$ to 10$^6$ images depending on the number of classes, image size and complexity, as well as network depth. Since the number of known strong lenses is smaller, it is necessary to create mock training sets.

\begin{figure}[htbp]
    \centering
    \includegraphics[width=0.98\textwidth]{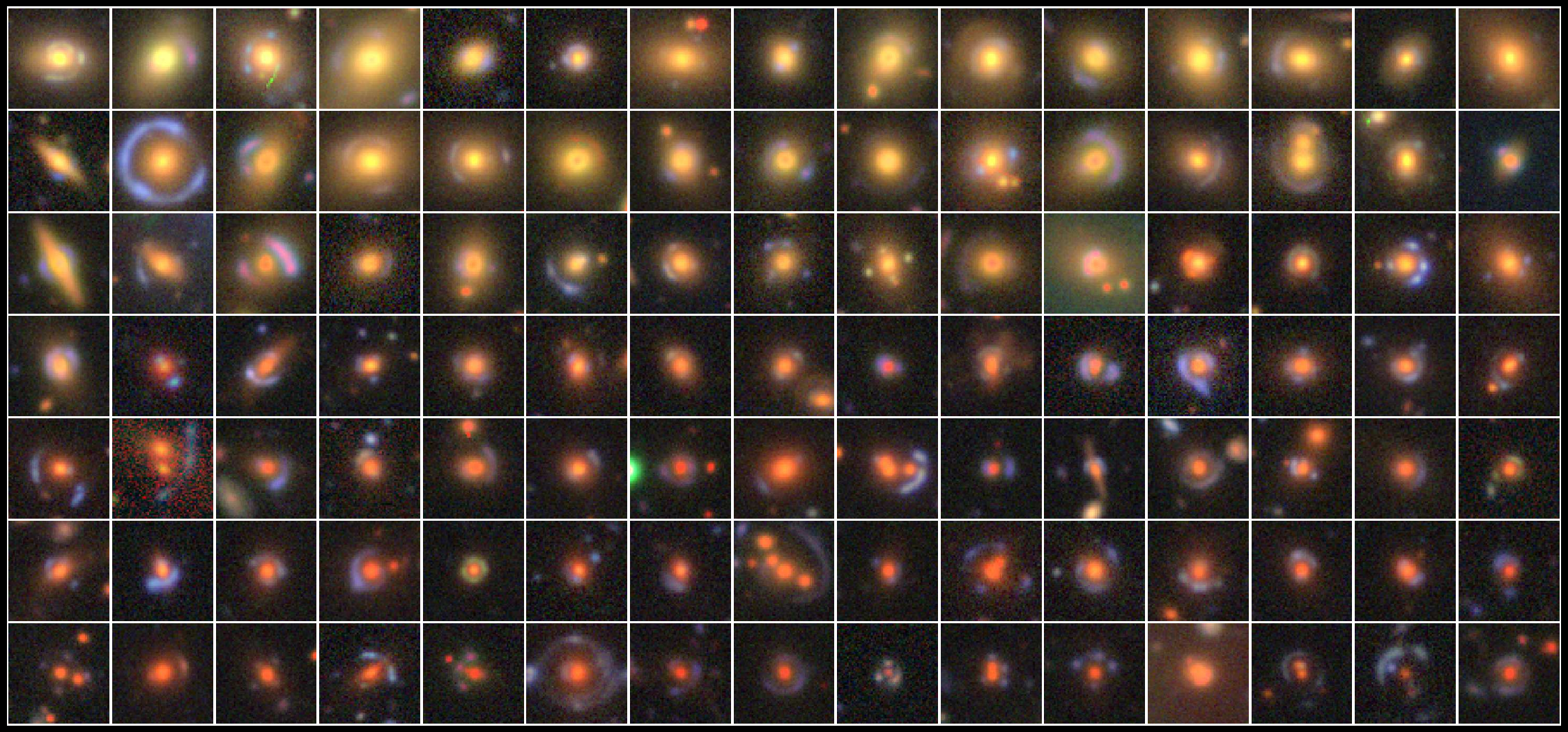}
    \caption{Colour composite images ($10^{\prime \prime} \times 10^{\prime \prime}$) of 105 high-quality strong lens candidates identified from the Hyper Suprime-Cam Subaru Strategic Program by \citet{Shu2022} using CNNs.}
    \label{fig:holismokes_mosaic}
\end{figure}

The success of pioneering searches \citep[e.g.,][]{Jacobs2017,Petrillo2017} and the results from the strong-lens classification challenge \citep{Metcalf2019} have demonstrated the efficiency of supervised CNNs trained on strong-lens simulations to identify galaxy-galaxy configurations. While improving the CNN architectures only leads to minor gains \citep[e.g.,][]{Schaefer2018}, generating highly-realistic lens simulations and training sets that account for the complexity of the actual survey datasets is the main ingredient to reach optimal classification performances \citep[e.g.,][]{Lanusse2018}. Building on these early studies, CNNs have been applied to a large range of single and multiband imaging surveys in the optical and near-infrared to select strong-lens candidates, including DES \citep{Jacobs2019,Jacobs2019b,Rojas2021}, KiDS \citep{Petrillo2019,Li2020,Li2021}, PanSTARRS \citep{Canameras2020}, HSC \citep[see Fig.~\ref{fig:holismokes_mosaic};][]{Canameras2021,Shu2022}, the DESI Legacy Imaging Surveys \citep{Huang2020,Huang2021,Storfer2022}, UNIONS-CFIS \citep{Savary2022}, and \textit{HST} fields \citep{pourrahmani2018}. In all cases, a visual inspection stage appears to be necessary for validating the CNN candidates and for increasing the sample purity by removing non-lens contaminants such as groups, spirals, and especially ring galaxies \citep[see][]{Rojas2021}.

Each of these searches has found tens to hundreds of high-quality candidates with apparent strong lensing features and, while the grading schemes vary between different teams$^{2}$\blfootnote{$^{2}$ Visual inspection by humans, including experts in strong lensing, also remains subjective \citep[see for instance Appendix A in][]{Shu2022} and \citet{rojas2023}.}, $\sim10^4$ new probable or definite galaxy-galaxy strong-lens candidates have been discovered with CNNs. Because confirming these candidates requires spectroscopic follow-up, ideally in combination with high-resolution imaging, the process remains highly time-consuming. 
%Confirmation and redshifts will come from the forthcoming fiber-fed spectroscopic survey facilities (e.g., 4MOST), and the wide-scale imaging surveys from space (e.g. Euclid, Roman).
Among the recent spectroscopic confirmation campaigns, \citet{Tran2022} have conducted a systematic follow-up of high-quality candidates from DES and DeCALS. The 88\% success rate they obtain provides evidence for the high purity of CNN-selected galaxy-scale strong-lens candidate samples. The use of CNNs for finding group- and cluster-scale lenses has been much more limited, mainly due to the higher difficulty in simulating the diversity of this population.

CNN-based strong lens classification studies follow the supervised approach, with binary classification (lens or not lens). Multi-class models have been shown not to offer a significant benefit in lens classification \citep{teimoorinia2020}. To the zero-th order, a supervised classifier can be regarded as a similarity checker. As a result, if the training set is biased in any specific way, decisions from such a classifier will be biased in the same way. For example, \citet{Shu2022} trained two strong-lens classifiers constructed from the same architecture with separate training sets that have different distributions in the lens galaxy redshift. Applied to the same set of strong-lens systems, it was found that, for both classifiers, the performance in lens galaxy redshift ranges that were under-represented in the training set was significantly worse than in lens galaxy redshift ranges that were well sampled. As illustrated with this example, the complex selection functions from supervised learning algorithms can fortunately be quantified with extensive tests based on real survey data \citep[see also][]{Canameras2021}. On the one hand, biases resulting from the design of training sets complicate the development of a self-consistent CNN classifier able to recover the variety of strong-lens morphologies, including the rare configurations \citep{Wilde2022}, while minimising confusion with non-lens contaminants. On the other hand, this sample bias can actually be exploited to design classifiers that target specific types of strong-lens systems. \citet{jacobs2022} quantified the sensitivity of lens-finding CNNs to various input properties in the training sample, finding the most important variables to include realistically are colour and PSF. \citet{herle2023} report that CNNs preferentially select systems with both larger Einstein radii and larger, more concentrated sources.

Overall, these limitations are encouraging the development of complementary data-driven approaches. Among these efforts, \citet{Cheng2020} have conducted unsupervised deep learning classification of \textit{Euclid}-like images, by combining a convolutional autoencoder with a Gaussian mixture model to cluster the extracted morphological features without the need for a training set, and \citet{Stein2022} have explored the capabilities of self-supervised representation learning using ground-based images. \citet{keerthivasan2022} have demonstrated the use of semi-supervised models in which the training sets are augmented with a Generative Adversarial Network (GAN), discovering $\sim$20 promising lens candidates in the deep, small-area survey, the Deep Lens Survey. \citet{thuruthipilly2022} demonstrated that self-attention-based encoders -- namely input-dependent weighting of features -- can outperform conventional CNNs in finding lenses in mock space- and ground-based imaging. These novel methods are promising, particularly for identifying exotic lens configurations that are difficult to include in training sets.

Perhaps surprisingly, CNNs have rarely been used to discover lensed quasars. This is likely due to current samples having faint lensing galaxies, leading to confusion between quasar-star projections and double lensed quasars when using optical imaging data alone. However, quadruply imaged quasars are uniquely identifiable when their image separation is large enough. Indeed \citet{akhazhanov2022} have shown that CNNs can reach 80\% classifcation performance on mixtures of mocks and known systems, and recently such a CNN-based search has discovered a new lens in Pan-STARRS \textit{gri} data alone \citep{dux2023b}. \citet{andika2023} find reduced contamination in real data when combining CNNs and vision transformers, and discover new candidate lensed quasars in HSC, as well as lensed galaxies as a fortuitous by-product.

\subsubsection{Decision trees} \label{machine_learning_catalogues}

%\chiara{Modern deep and wide-field sky photometric surveys offer an unprecedented opportunity to collect large amounts of data for statistical studies of the nearby and far-away universe. 
%One of the fields where large sky surveys can be of precious help is in finding new lensed quasars. In fact, strongly lensed quasars are very rare objects: one quasar in $\sim10^{3.5}$ is expected to be strongly lensed for an i-band limiting magnitude $\sim21$ mag (see Fig.~3 in \citealt{Oguri2010}).}

Decision trees are one of the oldest structures for building supervised learning models. They can be thought of as a sequence of if-else conditions, that visually take the form of a tree. An input passes through the tree and finishes with a classification. While passing through the tree, various attributes of the input decide which subsequent branches to take.
%Provided a learning set of observations, the learning algorithm will first browse every input attribute and all values within these attribute in order to select the one that best split the set of observations according to a chosen metric. The learning set of observations is then split and each subset is recursively processed in the same way until no more split is needed. 
The main advantage of decision trees is that they are easy to interpret, though they often provide poor performance given their sensitivity to the training set that is used. %would have provided us with a very different model and hence different predictions). 
They are said to have a high \emph{variance}. In order to mitigate against this effect, one can instead build $N$ training sets, compute $N$ trees and average all their predictions. This approach, known as Bootstrap aggregation, is used in practice in the Random Forest method along with a random selection of the input attributes. Another variant includes Extremely Randomised Trees (ERT), in which both the input attributes and observation splits are randomly selected, and many such trees are combined to provide predictions.

%Another approach resides in the random selection of both the input attributes and of the split values, the best split selected over a finite set of $K$ trials being then retained. The combination of the predictions of such $N$ so-called weak learners then provide performance that usually compete or even surpass those of more sophisticated approaches, like ANN, along with a significant faster learning phase. Extremely Randomised Trees (ERT) are based on this idea.

\sloppy
As such, decision tree algorithms perform well in classifying extragalactic sources, in particular quasars \citep[e.g.][]{ball06}. Working at the catalogue level, they allow the user to explore large datasets with little human intervention and affordable computing time, thus selecting candidates with less stringent pre-selection criteria, maximising the precision (recovery rate), and minimising the stellar contamination. This was used as a way to remove stellar contamination from the lensed quasar search of \citet{khramtsov2019}.

As already mentioned in Section \ref{catalogue_selection}, quadruply-imaged quasars occupy specific regions of image-position and flux-ratio space. Large astrometric catalogues of individual quasar positions thus constitute an opportunity to detect lensed quasars through the use of decision trees, i.e. without having to rely on complexities of processing astronomical images. \textit{Gaia} provides an all-sky high-resolution catalogue of quasars, ideally suited to the search of strongly lensed quasars. Using an ERT on {\it Gaia} DR2 data, \cite{Delchambre2019} achieved an identification rate of quadruply-imaged quasars that is better than 90\% along with a misclassification rate of stellar asterisms below 1\%. Nevertheless, it should be noted that the miss of a single image leads to a misidentification rate between 10 and 20\%, depending on the lensed image that is unobserved, due to the dimensionality reduction and associated loss of constraints.

%%%%
\section{Selection biases} \label{sec5:bias}% and Statistical applications [Paul]} \label{sec5:bias}
Having described various lens searches, we now turn to the possible biases from these searches. Lenses are often used to probe the source or lens population, and require mass modelling. The priors that apply to these models should not be uniform because of the way that the lens system was selected. For example, magnitude-limited lens samples are biased towards higher source magnifications, and thus the sources do not lie uniformly in the source plane, but rather they preferentially lie closer to the high-magnification caustics. This might not strongly bias results when the data are strongly constraining, however population level studies can fall foul to neglecting selection effects. It is important for anyone using lenses to keep in mind the various biases of existing samples of lenses. Furthermore, understanding the limitations of previous selections might initiate novel searches aimed at removing any harmful biases from these samples. 

\citet{mandelbaum09} provide an in-depth exploration of the selection effects of lensed point sources using a mock generation pipeline with mass models composed of both stellar and dark matter components. \citet{sonnenfeld23} extend this investigation to extended sources paying particular attention to stellar and dark matter masses, lens half-mass radius, and source size and surface brightness. Below, we briefly summarise several biases that exist in strong lens samples and explain their origin.

\begin{itemize}
\item{\bf Magnification bias:}
For a magnitude-limited sample, all of the intrinsically faint sources in the sample must have large magnifications. This leads to biases towards sources near caustics (by definition the highest magnification regions), and thus with image configurations corresponding to cusps and folds. For the same reason, lenses with larger caustics are more prevalent, which are those with more elliptical potentials or with multiple lensing galaxies (see also the quad ellipticity-shear bias this causes, described later).
%All quasars are at z=2; if you see a very bright lensed quasar, it's not because it's closer, it's because it is more highly magnified.

\item{\bf Redshift bias:}
Spectroscopic searches are limited to requiring multiple emission lines within the spectral range. The majority of spectroscopic searches are performed in the optical, and thus are limited to only one strong emission line above $z\approx1$. Furthermore, photometric selection of sources, such as quasars, has a redshift-dependent completeness. Above $z=2.7$, the UV-excess used to identify quasars is no longer present due to the $u$-band dropout of Lyman alpha, so more stringent constraints are used to remove stellar contamination, leading to less complete quasar discovery. A redshift bias can also manifest through a bias towards brighter lensing galaxies: as lensed quasar candidates with brighter lensing galaxies (as seen in existing optical imaging) are preferentially selected for limited spectroscopic confirmation, those systems with higher redshift (and thus fainter) lenses are not confirmed. Lower redshift sources naturally have lower redshift lensing galaxies, and are thus brighter and more likely to be confirmed. This is demonstrated in the known lensed quasar sample in Figure \ref{fig:completeness}.
%sources at some redshifts are more easily confirmed than sources at other redshifts.

\begin{figure*}[!htb]
\centering
\includegraphics[width=\textwidth]{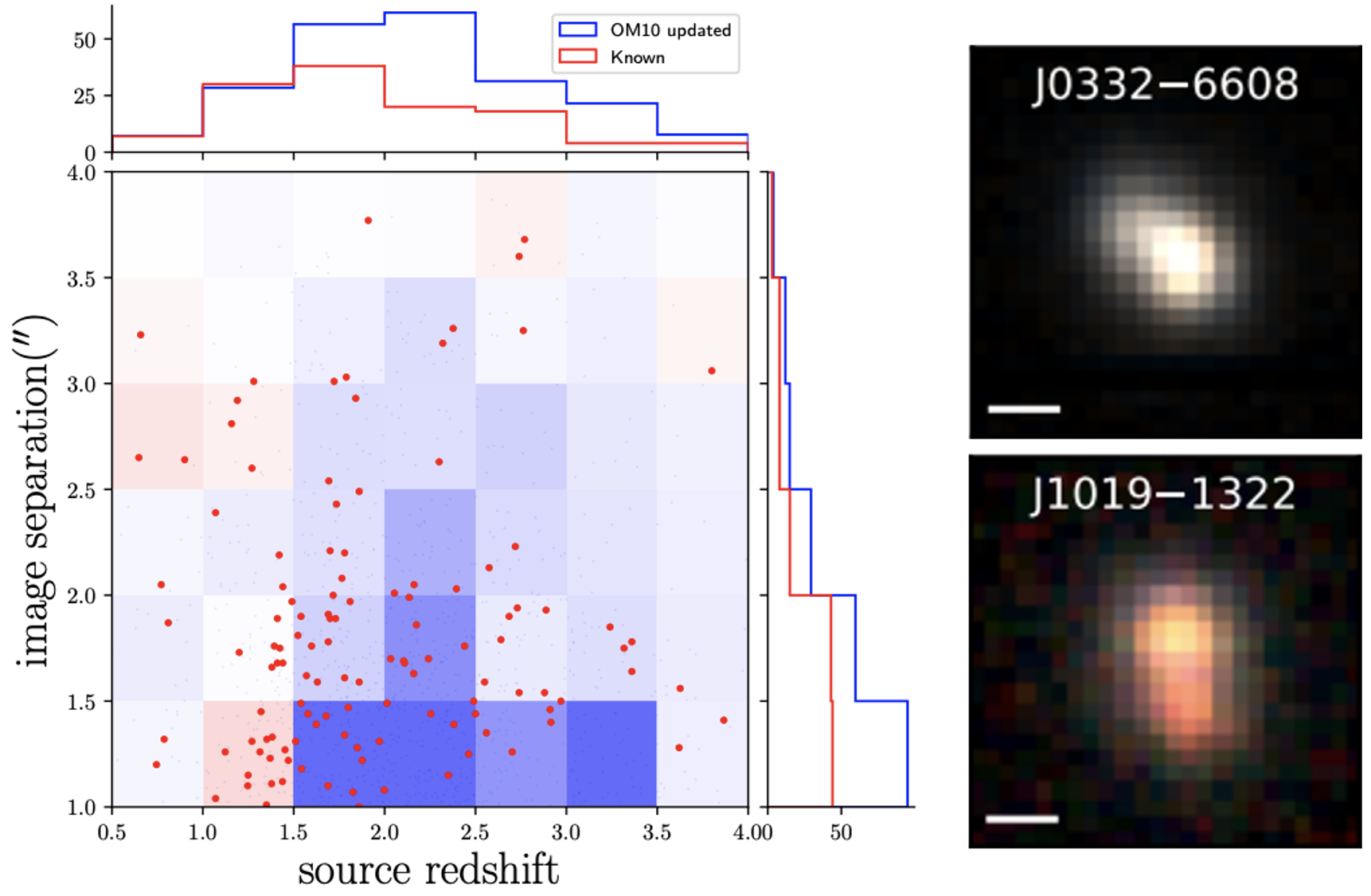}
\caption{\textit{Left}: Distribution of image separations against source redshifts for known lensed quasars (red) satisfying certain `discoverable' criteria, overlaid on an updated mock catalogue of \citet{Oguri2010}. There is a clear lack of known lenses (regions shaded blue) with smaller separations and higher redshifts. \textit{Right}: examples of quasar pairs with similar spectra from \citet{lemon2023} that may represent these missing systems, lacking only a lens galaxy detection.}
\label{fig:completeness}    
\end{figure*}

\item{\bf Einstein radius bias:}
Many search techniques rely on resolving the multiple images of a background source, however if these images are unresolved at the catalogue and/or pixel level then the lens cannot be selected in the original parent sample. This leads to a lower limit on the lensing galaxy Einstein radii. Conversely, fibre-selection of lens candidates (see Section \ref{fibrespectra}) can miss those systems in which the brightest images lie outside the fibres, i.e. when the Einstein radii are larger than the fibre radius.

%The resolution of a survey is a 
%Lenses must have large masses to produce observable separations between multiple missions

\item{\bf Model bias:}
The mass models that are used to create mock lenses do not capture the complexity of real lens systems, and thus both humans and machines trained on such mocks are at risk of missing real lenses with non-standard mass distributions. %without similar mass models. %A poor model for fitting a real lens also leads to biases in the inferred parameters.

%bad model bias: failure to correctly model a system can lead to systematic errors in the inferred parameters.

%\item{\bf Visual inspection bias:}
%Lens searches often end in the visual inspection of thousands of candidiates selected by some algorithm A final catalogue of high-confidence lenses are usually subject to visual inspection

\item{\bf Microlensing bias:}
Lens masses with Einstein radii comparable to the source size, such as individual stars and quasar accretion disks, can cause considerable magnifications and demagnifications of the background source. This effect depends on the convergence, internal shear, and stellar mass density at the image positions, and is more dramatic for saddle points \citep{schechter02}. These discrepant flux ratios can manifest as `missing' images in shallow imaging (see Figure \ref{microlensing_0924}), which can lead to systems being missed by various magnitude-dependent selection techniques.

%Stage1: 
%Microlensing may cause stars in lensed system to be brighter; lenses with large stellar densities may be preferentially included in a sample.
%Stage2: 
%Microlensing may cause stars in lensed systems, in particular the 2nd and 4th brightest in doubles and quads, respectively, to be fainter.  Lenses with large stellar densities may be preferentially excluded from a catalog.

\begin{figure*}[htb]
\centering
\includegraphics[width=0.85\textwidth,angle=0]{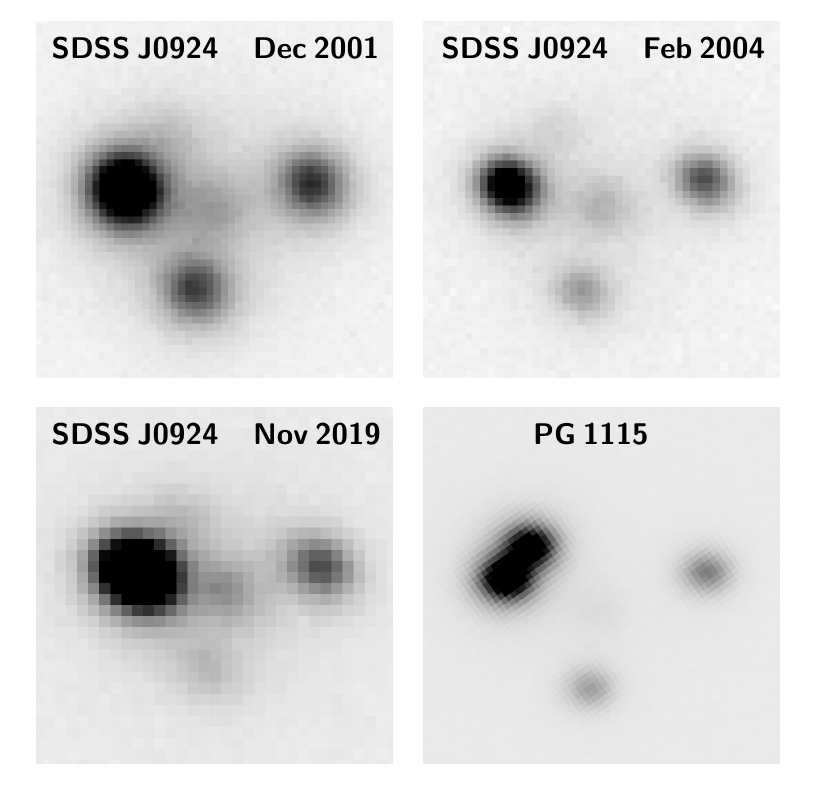}
\caption{$i$-band images of two quadruply imaged lensed quasars: SDSSJ0924+0219 shows one image being severely demagnified by microlensing over two decades, and PG1115+080 during an epoch without strong microlensing. The top two images were taken with MagIC, and the bottom two with IMACS. We note that PG1115+080 has a larger Einstein radius, and thus a larger cutout.}
\label{microlensing_0924}
\end{figure*}

\item{\bf Variability bias:} \label{variabilitybias}
\citet{lemon2020} uncovered a \textit{variability bias} associated with finding lensed quasars through variability alone. This bias is due to a well-known correlation between variability amplitude and quasar luminosity \citep[see, e.g.,][]{macleod2012}, resulting in the lenses with the faintest sources having the most variable images. Quads have, on average, higher magnifications than doubles, and so for the image brightness, quads will vary more since they probe fainter sources.

\item{\bf Survey length bias:}
\citet{oguri2003} pointed out that cadenced surveys for detecting lensed supernovae will miss images and lens systems as the duration of the survey approaches the time delay, leading to biases towards both more symmetric and smaller Einstein radius systems.

\item{\bf Time-delay selection function} 
Similar to the Survey length bias above, constraints on observations -- both spatially or temporally -- prohibit the discovery of lensed transients found in the time domain through the reappearance of extra images. For example, FRBs are of millisecond duration, and many widefield instruments (e.g., CHIME/FRB) are incapable of slewing. This imposes a ``time-delay selection function''~\citep{connor2022stellar}, which selects for strongly-lensed FRB systems whose time delays are either (i) less than the amount of time spent by one sky position within the beam pattern, or (ii) an integer multiple of the sidereal day. 

\item{\bf Caustic area bias:}  
\citet{baldwin21} explore the Malmquist-like bias of quadruple image lensed quasars having larger caustic areas. Models that do not account for this selection effect typically have caustic areas that are biased low, and as such the predicted time delays are also biased low, leading to an important bias in the Hubble constant from time-delay cosmography of the order of $\sim$1\%.

%\item{\bf Mass sheet bias:} 
%Sources with bigger mass sheets have bigger caustics

%\item{\bf Circularity bias}  ADDED TO NEXT BIAS
%Near-circular lenses, particularly for near-isothermal lenses, produce higher magnifications than flattened lenses; but they also have smaller caustic areas; the {\it sign} of this bias is unknown.

\item{\bf Quad ellipticity-shear bias:} Quadruple image lenses are biased towards more flattened lensing potentials, due to more elliptical lensing galaxy masses and/or larger external shears. This effect is less strong for extended sources than point sources, as sources can often overlap a large fraction of the background caustics.  Conversely, very circular lenses have smaller caustic areas and are thus less likely to be discovered as bright quads, though this might be compensated for by the higher magnifications associated with near-circular lenses.

\end{itemize}

\section{Prospects for future lens searches} \label{sec6:future}
%CS: There might be a new chapter covering this. This section should only introduce new missions/facilities but without giving technical specs. (CS 25/10/2022. Following the ISSI plenary). multi-wavelength searches, LVEK: Euclid, SKA, Sub-mm ?

As emphasised throughout this Chapter, strong gravitational lensing is a rare phenomenon that spans a very broad range in lens and source properties. It is thus important that very different techniques are used to discover strong lenses. Our playground for this exercise resides in the many multi-wavelength, time-domain and even multi-messenger datasets and surveys that are ongoing or planned in the very near future. We outline which future surveys and observatories will provide the drive for finding orders of magnitudes more lenses. The section is divided by source type, though there is naturally significant overlap between surveys providing the relevant data for finding each type of lens.

\subsection{Lensed galaxies} \label{future_galaxies}

The ESA-NASA \textit{Euclid} survey will image 15,000 square degrees of extragalactic sky in one wide optical filter (550--920nm) and three near-IR filters: Y, J, and H \citep{Laureijs2011, Scaramella2022}. The first strength of \textit{Euclid} will be its spatial resolution of 0.18\arcsec, giving access to the most compact systems with Einstein radii down to 0.3-0.5\arcsec. The sharp \textit{Euclid} PSF increases the contrast between the smooth lens light and the thin arc-like images of the background source. While the broad band-pass of the \textit{Euclid} optical filter will optimise the detection of lensed sources with strong continuum emission, it may be less effective at detecting lensed sources with flux dominated by strong emission lines. Other future space-based surveys like the Roman space telescope or the Chinese Space Station Telescope (CSST) will complement the capabilities of \textit{Euclid} through increased depth and number of filters. 

\citet{collett2015} create mock \textit{Euclid} lens observations to estimate the number of discoverable systems, assuming at least partially resolved sources, magnifications above 3, and sufficient signal-to-noise. If the lens can be well subtracted, 170,000 lenses should exist in \textit{Euclid} with a peak Einstein radius around 0.5\arcsec. For comparison, an existing survey like DES is predicted to contain 2400 lenses with peak Einstein radii around 1.25\arcsec.

%In this case, the lines may be diluted in the shot noise from the sky background and other astronomical objects (including the lens). The near-IR camera onboard \textit{Euclid} should, however, allow to circumvent this effect by enlarging the wavelength coverage in regions where the lens and source differ the most. 

The Rubin-LSST survey will image the whole Southern sky in 6 optical filters every few days. This adds the time and colour dimensions to \textit{Euclid} and will surpass the depth of \textit{Euclid} after a few months of observation but, as a ground-based survey, will be limited by atmospheric turbulence. It will play a major role in the colour pre-selection of targets, providing photometric redshifts, and in the lens-source deblending using machine learning or morpho-spectral techniques \cite[e.g.][]{Joseph2016, Melchior2018}. \citet{collett2015} also predict the number of lenses to be discovered in LSST data alone: 120,000 in the optimal stack, and if the lensing galaxy must be removed through a blue-red band difference imaging technique, then 62,000. The peak of the Einstein radii is around 1\arcsec, and source redshifts around $z=2$. % and perturb lens models. Combining the strengths of space-based and ground-based surveys is clearly the path to follow to enable efficient searches for strongly-lensed systems. 

The depth and coverage of both LSST and Euclid will expand galaxy cluster catalogues to higher redshifts, through simple conventional searches of overdensities in catalogued nearby galaxies with similar colours and magnitudes \citep[e.g.,][]{gladders2005}. The ongoing eROSITA X-ray survey will also play a role in finding $\sim$ 100,000 X-ray-bright clusters out to $z>1$. The discovery of giant arcs in many of these clusters will be possible with the discovery datasets themselves.

%TALK ABOUT SUB-MM, JWST, SKA, ngVLA
At sub-mm wavelengths, the models that correctly reproduce the dominance of lensed systems in sub-mm galaxies (SMGs) at flux densities above 100 mJy, predict that $>1$\% of systems down to 1 mJy should be lensed -- 10 lensed SMGs per square degree -- in agreement with the number of lenses found in a common JWST-SCUBA-2 imaging field \citep{pearson2023}. Discovery and confirmation of these systems is not possible at optical wavelengths, but could be possible in the wide-field infrared imaging of \textit{Euclid}. JWST, ALMA, and the ngVLA \citep[next-generation Very Large Array, to be fully operational by 2034,][]{selina2018} will be key to demonstrating the efficiency of these searches, and providing higher quality datasets to study the source and lens populations.

\subsection{Lensed quasars}
Lensed quasar searches have relied predominantly on source selection, i.e. starting from a catalogue of known quasars and high-confidence quasar candidates. These have come from spectroscopic datasets, such as the 2dF Quasar Redshift Survey (3$\times$10$^{4}$), SDSS/BOSS (5$\times$10$^{5}$), as well as photometric selections due to UV-excess or typical red colours in the infrared (e.g. 10$^{6}$ from \citealt{secrest2015}). These previous searches will soon be extended to fainter magnitudes, as an order of magnitude more quasars are expected from upcoming spectroscopic surveys in the next 5 years, such as the Dark Energy Spectroscopic Instrument (DESI; 4$\times$10$^{6}$) and the 4-metre Multi-Object Spectroscopic Telescope (4MOST) extragalactic surveys (3$\times$10$^{6}$). Pushing to even deeper magnitude limits, the Maunakea Spectroscopic Explorer (MSE) is planned on an upgraded version of the Canada France Hawaii Telescope (CFHT) with a 11m mirror, and the capability of simultaneously obtaining 4000 spectra. Photometric selection of $z<2.7$ quasars will also be extended thanks to the $u$-band imaging of the Rubin-LSST survey in the South.

\begin{figure*}[htb]
\centering
\includegraphics[width=\textwidth,angle=0]{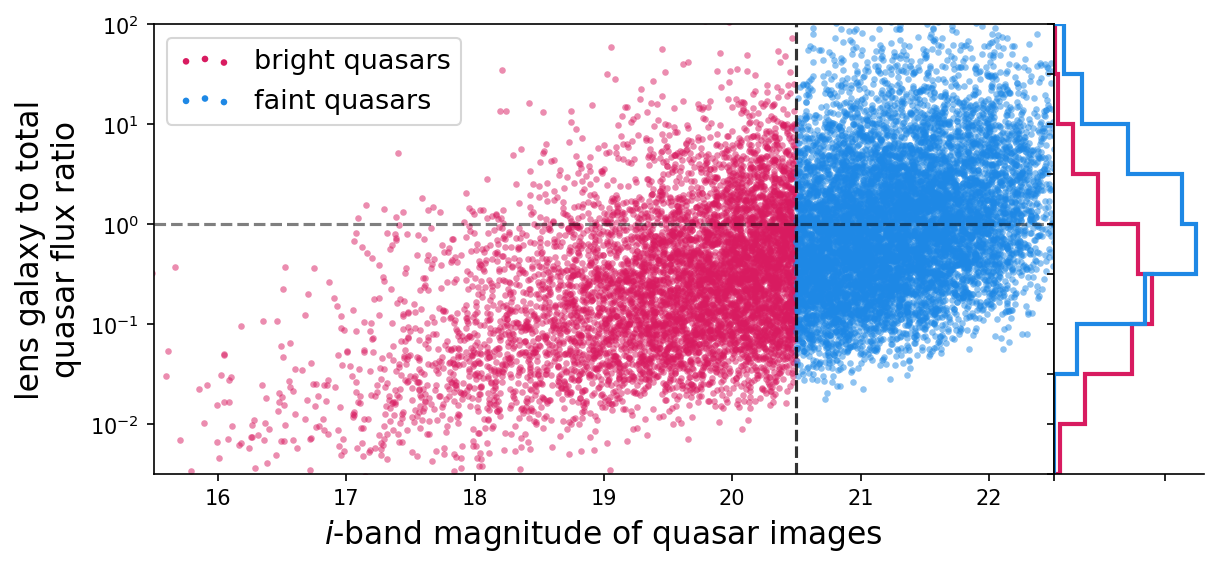}
\caption{Distribution of the lens-to-quasar flux ratio against total quasar magnitude for the mock lensed quasars of \citet{Oguri2010}. The lensing galaxy population is naturally the same at all source brightnesses, so the galaxy flux (and hence colours) will start to dominate the total system flux at the fainter quasar source magnitudes expected to be discovered in future surveys like LSST and \textit{Euclid}. }
\label{fig:lens_qso_brightnesses}    
\end{figure*}

However, at fainter magnitudes, lensed quasar searches can no longer be source-selected \textit{at optical or infrared wavelengths}. As demonstrated in Figure \ref{fig:lens_qso_brightnesses}, for lensed quasars with total image brightnesses fainter than 21, the galaxy light begins to dominate. In such cases, the selection must rely on galaxies, as spectroscopic and photometric quasar catalogues will miss lensed quasars due to the lens light contamination. Lens-selected searches are already the standard for lensed galaxy searches, and should readily be applicable to quasar sources as well; for example, a SLACS-style selection to search for quasar emission lines in galaxy spectra (Section \ref{fibrespectra}). In the X-ray and radio, there is often no contamination from the lens, offering a promising initial source-selected catalogue containing lensed quasars. The ongoing eROSITA X-ray survey will provide a deep, full-sky map at 2-10 keV at 15\arcsec\ resolution, expected to detect 3$\times$10$^{6}$ AGN. Cross-matching optical catalogues to X-ray sources has seen success for lensed quasar discovery \citep[e.g., between HSC and \textit{Chandra,}][]{jaelani2021}, and even already between eROSITA and \textit{Gaia} \citep{tubin2023}. %At radio wavelengths, ??.

Another promising selection uncontaminated by the brightness of the lensing galaxy is through extended variability (see Section \ref{variability}). Briefly, through multiple epochs of the same field in the same filter, all lensed quasars and quasar pairs may be found through \textit{difference imaging}. LSST will provide the ideal dataset for this search: an approximate cadence of 3 days over 20,000 square degrees in 6 filters. By stacking epochs over the 10 year nominal lifetime, variability should be detectable down to $r\sim26.5$, or deeper when using multiple bands for detection. \citet{taak2023} predict $\sim$1000 lensed quasars to have variability detectable within LSST.
%The common contaminants of lensed quasar searches, such as star-forming galaxies and quasar+star projections, will appear as non-variable, or as single point source variables. Any extended variables are likely lensed quasars or the still physically rare and interesting quasar pairs and projections. 
%We also note a highly pure all-sky catalogue of the brightest quasars will soon be constructed through \textit{Gaia} data alone (7$\times$10$^{5}$), thanks to the increased baseline, providing more precise proper motions, parallaxes, and variability information. 

We can also expect a push towards the discovery of smaller separation lensed quasars. High-resolution imaging of a handful of radio-loud or optically bright quasars has led to the smallest separation known lenses: only four systems confirmed below 0.4$^{\prime\prime}$. While \textit{Gaia} offers a 0.2$^{\prime\prime}$-resolution map of point sources across the whole sky, cataloguing issues in existing data releases at small separations and in crowded local fields leads to the loss of several detections. If such issues are resolved in future releases, or the 1D single-epoch windows are provided, 100s of predicted, bright small-separation lenses might be able to be confirmed within the survey data themselves. At fainter magnitudes, \textit{Euclid}'s space-based resolution and wide area (see Section \ref{future_galaxies}) will yield many new lens systems not only from analysis of known quasars and quasar candidates, but will also detect the lensing galaxies of many existing unclassified quasar pairs, for which ground-based imaging is too shallow to detect any lenses.

At radio wavelengths, the International LOFAR Telescope (ILT) will observe 15,000 square degrees at $\sim$150 MHz with 0.35\arcsec\ resolution and a 1$\sigma$ sensitivity of 90$\mu$Jy. \citet{rezaei2022} show that CNNs can recover 95\%  of lensed radio sources with Einstein radii above 0.5\arcsec\ and flux densities above 2 mJy, while removing nearly all contamination from unlensed double-lobed sources. At similar resolution, a deeper all-sky survey with the Square Kilometre Array (SKA) could uncover 10,000 new radio-loud lenses, with sources mixed between lensed radio jets and lensed starbursts, with source redshifts potentially at $z>10$ \citep{mckean2015}. Though the exact surveys of SKA are not yet confirmed, observations are planned to begin around 2028/2029.

\subsection{Lensed supernovae}
The planned Legacy Survey of Space and Time (LSST), to be carried out with the Simonyi Telescope of the Vera C. Rubin Observatory, promises an unprecedented dataset for finding lensed supernovae, thanks to its high-cadence, wide-field, and deep imaging. 
As discussed in Section \ref{sec4:seclection_methods} there are both catalogue-based schemes and pixel-based schemes for discovering lensed supernovae. The simplest of catalogue-based schemes are based on the assumption that two or more images of the lensed supernovae are recorded as transients. Given a conservative estimate of a detection limit of 22.6 mag in the $i$-band, an effective survey length of 2.5 years, and image separations above 0.5\arcsec, \citet{Oguri2010} predict a total of 130 galaxy-scale lensed supernovae from LSST. Unresolved catalogue schemes based on strong magnifications of type Ia supernovae from measuring redshifts either photometrically or spectroscopically (see Section \ref{physical_relations}) are also possible, and \citet{goldstein2019} predict 380 lensed supernovae in LSST (10 years) with such a method, broadly agreeing with the 180 lensed type Ia supernova estimate of \citet{sainzdemurieta2023}. \citet{wojtak2019} consider both discovery methods, finding a much higher rate of 340 lensed supernovae \textit{per year}, with $\sim$2/3 of these discoverable through their magnification alone. Combining multiple epoch observations and a variety of selection methods, such as CNN classification \citep[e.g.,][]{kodiramanah2022}, will be key to discovery of these systems. The delivered image quality, cadence, observing strategy, and access to follow-up facilities will all be important factors for the true discovery rate \citep{huber2019}. 

\begin{figure*}[htb]
\centering
\includegraphics[width=\textwidth,angle=0]{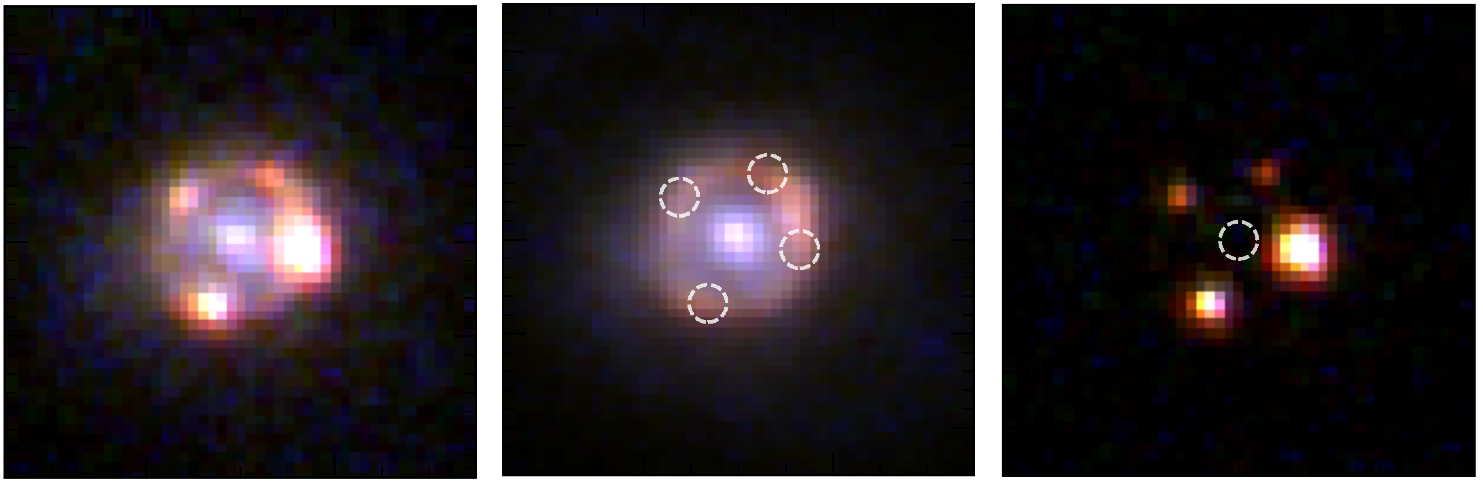}
\caption{Left: \textit{HST} observation of the lensed supernova iPTF16geu; \textit{middle}: observation after the supernova has faded; \textit{right}: subtraction of the constant components, showing just the supernova images. Reproduced from \citet{dhawan2020}.}
\label{fig:iPTF16geu}    
\end{figure*}

%include! ana's results
As LSST will also provide deep existing imaging of any candidate, and, in the absence of lens confirmation through spectroscopy, deconvolution schemes \citep[e.g.,][]{akhaury2022} may be used to detect a lensed host galaxy before possible further images arrive. In a similar fashion one might look for a leading image that was too faint to catalogue -- ``pre-covery'' -- as the source transient might fall near a fold caustic or due to extrinsic microlensing, similar to the case of iPTF16geu (see Figure \ref{fig:iPTF16geu}).

The number of known lensed galaxies, detected through large spectroscopic surveys or through arc detection in imaging (see Section \ref{future_galaxies}), will increase by an order of magnitude before the end of the lifetime of LSST. Cross-matching known transients to lens candidates \citep{magee2023, sheu2023} and monitoring of known lenses \citep{craig2021}, will become a fruitful method for lensed supernova discovery, as it already has been for monitoring of cluster lenses \citep[e.g.,][]{kelly2015, rodney2021}. JWST will become a key tool for lensed supernova discovery in galaxy clusters given its unprecedented depth and resolution and thus ability to reach higher redshift fainter supernovae. Indeed, \citet{frye2023} already report three images of a new lensed supernova in galaxy cluster G165. Based on star formation rates of source galaxies, they suggest as many as 1 more lensed supernova to be expected from monitoring of that one cluster alone. \citet{shu2018} predict over 10 lensed supernovae occur per year in a sample of 128 known lensed galaxies from various SDSS lensed galaxy searches. 

Such monitoring of clusters with the \textit{HST} has also recently revealed individual lensed stars and star clusters \citep{miraldaescude1991, kelly2018} thanks to the magnifications (both macro and micro) near critical curves of factors of several thousand. \textit{JWST} is already delivering new systems \citep[e.g.,][]{chen2022b}, and will be the most powerful observatory for the discovery of such systems for the foreseen future.

 %that of SN Refsdal \citep{kelly2015}, discovered with the Hubble Space Telescope (HST) and lensed by a cluster of galaxies, iPTF16geu \citep{goobar2017}, discovered with the interim Palomar Transient Faciltiy and lensed by a single galaxy.  The latter was unresolved in its discovery imaging, but was subsequently observed extensively with the HST \citep{}, as shown in Figure \ref{fig:iPTF16geu}.

\subsection{Lensed GWs}
As LIGO--Virgo--Kagra (LVK) currently enter their fourth observing run (O4) at design sensitivity, detecting lensed GWs from stellar-mass black hole (BH) mergers becomes a possibility, and very likely in the fifth observing run (O5) sensitivity (estimated to start in 2027). Predictions for lensed GWs suggest rates around 0.1 -- 5 lensed BH-BH mergers per year in O4, and 1 -- 50 per year in O5, with the range caused by our uncertainty on underlying compact merger rates \citep[e.g.,][]{li2018, ng2018}. Lensed NS-NS events are unlikely to be detected, even during O5 \citep{smith2023}. Most time delays will be shorter than one month, and $\sim$30\% of lensed events are expected to be quadruply imaged \citep{li2018}.

Several next-generation GW observatories (so-called third-generation) are currently planned and will be sensitive to lensed GWs of various origins. Many of these observatories will also have superior angular resolutions of 1\arcsec or better, and extend the detection volume out to high-redshift, offering the chances of detecting optical counterparts of lensed NS-NS mergers. The Einstein telescope will have sensitivity to NS-NS events out to $z=2$, and thus even higher redshift for lensed events. \citet{piorkowska2013} predict up to several tens of detections per year, and an order of magnitude more lensed BH-BH mergers \citep{biesiada2014}. 

The Laser Interferometer Space Antenna (LISA) is a planned (2030s) space-based set of spacecraft that will be separated by 8 light-seconds, and will provide sensitivity to larger volumes, at lower frequencies, and accordingly higher total merger masses. Optimistic models suggest up to 4 well-detected strongly lensed events within the 5-year nominal mission \citep{sereno2010}. Between the LISA-LIGO frequency gap, DECIGO, a planned space-based mission, will operate at deci-hertz frequencies. The principle mission is to detect primordial GWs expected from inflation, however it will also detect Intermediate Mass Black Hole mergers and their lensed counterparts \citep{piorkowskakurpas2021}.

\begin{figure*}[htb]
\centering
\includegraphics[width=0.85\textwidth,angle=0]{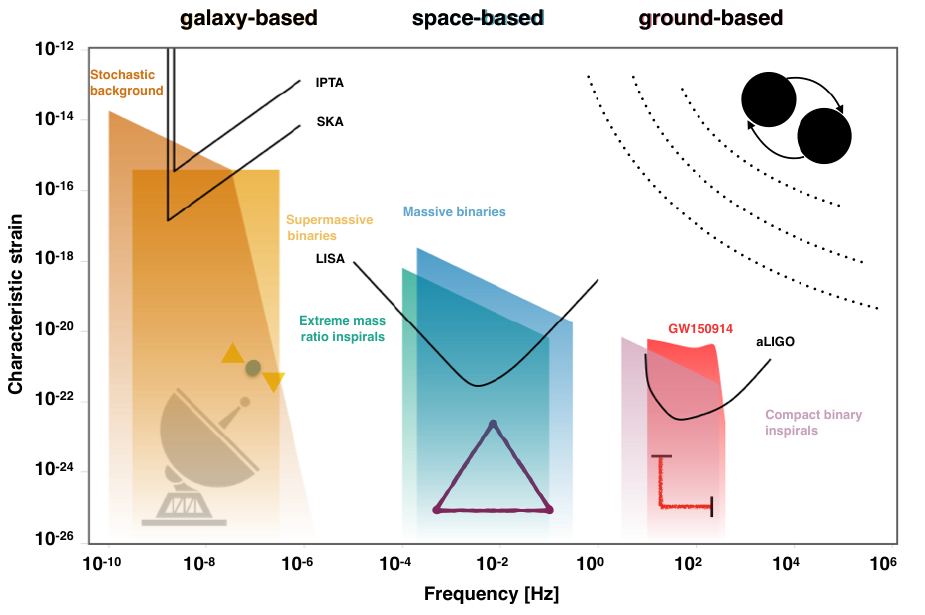}
\caption{ Gravitational wave spectrum. It shows the frequencies at which various GW observatories (current and upcoming) are going to operate and correspondingly astrophysical sources. Figure is reproduced from K. Islo's Thesis. Credit: C. Moore, R. Cole, C. Berry and C. F. Mingarelli.}
\label{fig:gwspec}    
\end{figure*}

At even lower frequencies, we expect to see GW signals from supermassive BH mergers. Pulsar Timing Arrays (PTAs) have recently reported strong evidence for a stochastic background consistent with such a population of BH mergers \citep{nanograv2023}. \citet{khusid2022} predict 10-30 lensed binaries might be detected using PTAs with the Square Kilometre Array. Tantalising evidence at optical wavelengths for such a system already exists in a known lensed quasar \citep{millon2022}. Figure \ref{fig:gwspec} summarises the frequencies and astrophysical sources of next-generation GW observatories. The sky localisation uncertainties depend on the signal-to-noise of the signal but typical localisation areas for the observatories (in square degrees) are $\sim$1 \citep[LISA,][]{kocsis2008}, $\sim$10 (advanced LIGO), and $\sim$40 \citep[SKA and iPTA,][]{sesana2010, finn2010}.

%The lensing rate predictions for advanced LIGO--Virgo--Kagra (LVK) and for the third generation (3G) ground-based era (including the Cosmic Explorer and Einstein Telescope, with observations aimed to start by 2035) indicate that we will have to wait for the first lensed GW sources in the 3G era, unless we are lucky enough to discover some extreme magnification events in the next few years with LVK. The LVK and 3G detectors operate upwards of 1~Hz wherein we catch mergers of low-mass neutron stars to stellar-mass BBH systems. 

%find signals from many inspiraling super-massive black hole (SMBH) binaries whereas LISA in the milli-hertz range will detect primarily their final stages of coaelescence in addition to detecting inspirals of intermediate-mass black hole (IMBH) binaries. 

%DECIGO, a planned space-based mission, will operate at deci-hertz frequencies filling the gap between LISA and the 3G detectors. While the main goal of DECIGO will be to detect primordial GWs expected to be produced during inflation, it will also detect the last stages of Intermediate Mass Black Hole (IMBH) mergers. Not only will these upcoming GW observatories have improved sensitivities but also superior angular resolutions of 1\arcsec or better. The science with lensed GW sources will give us unprecedented insights into the black hole population formation channels and their mass functions and similarly, insights into the physical processes responsible for kilonovae, the pre-merger states of neutron stars, and objects in the mass gaps.

\subsection{Lensed FRBs}

Perhaps the most promising path to detecting galaxy- or cluster-lensed FRBs involves a combination of microlensing searches and high-precision FRB localisations to their host galaxies. In this scenario, optical follow-up can confirm the lensed host galaxy. This will be enabled by widefield FRB surveys such as CHIME/FRB Outriggers~\citep{leung2021synoptic,mena2022clock}, CHORD, the DSA-2000, or future coherent all-sky monitors such as BURSTT~\citep{lin2022burstt}, which will overcome the time-delay selection bias by extending the detection field-of-view down to the horizon.
%{\bf In the radio, LOFAR and later-on SKA1 and SKA2} will extend this to ..., including lensing of FRBs... 

\begin{acknowledgements}
We thank the anonymous referees for their thorough reading of the manuscript and the feedback that has significantly improved the review. We thank the International Space Science Institute in Bern (ISSI) for their hospitality and the conveners for organizing the stimulating workshop on $``$Strong Gravitational Lensing$"$. This project has received funding from the European Research Council (ERC) under the European Union's Horizon 2020 research and innovation programme (LENSNOVA: grant agreement No 771776).
\end{acknowledgements}

\noindent\small{\textbf{Conflicts of interest}}
The authors have no relevant financial or non-financial interests to disclose. The authors have no conflicts of interest to declare that are relevant to the content of this article. All authors certify that they have no affiliations with or involvement in any organization or entity with any financial interest or non-financial interest in the subject matter or materials discussed in this manuscript. The authors have no financial or proprietary interests in any material discussed in this article.

\bibliographystyle{aasjournal}   
\bibliography{references}                % name your BibTeX data base
\nocite{*}

\end{document}